\documentclass[fleqn,usenatbib]{mnras}
\let\oldhref\href
\renewcommand{\href}[2]{\oldhref{#1}{\hbox{#2}}}

\usepackage{supertabular}
\usepackage{graphicx}	
\usepackage{amsmath}	
\usepackage{amssymb}	
\usepackage{multicol}        
\usepackage{bm}		

\defcitealias{lehner11}{LH11}
\defcitealias{lehner12}{L12}
\defcitealias{marasco22}{Paper II}
\defcitealias{marasco19}{M19}

\def\nodata{ ~$\cdots$~ }

\newcommand{\mnhi}{N_{\rm H\,I}}
\newcommand{\mnhii}{N_{\rm H\,II}}

\newcommand{\mlnhi}{\log N_{\rm H\,I}}

\newcommand{\km}{${\rm km\,s}^{-1}$}

\newcommand{\hst}{{\em HST}}

\newcommand{\vlsr}{\ensuremath{v_{\rm LSR}}}

\newcommand{\hi}{H$\;${\small\rm I}\relax}

\newcommand{\alii}{Al$\;${\small\rm II}\relax}

\newcommand{\nni}{N$\;${\small\rm I}\relax}
\newcommand{\nai}{Na$\;${\small\rm I}\relax}
\newcommand{\caii}{Ca$\;${\small\rm II}\relax}
\newcommand{\caiii}{Ca$\;${\small\rm III}\relax}

\newcommand{\cii}{C$\;${\small\rm II}\relax}

\newcommand{\civ}{C$\;${\small\rm IV}\relax}

\newcommand{\oi}{O$\;${\small\rm I}\relax}

\newcommand{\sii}{S$\;${\small\rm II}\relax}
\newcommand{\siii}{Si$\;${\small\rm II}\relax}
\newcommand{\siiii}{Si$\;${\small\rm III}\relax}

\newcommand{\siiv}{Si$\;${\small\rm IV}\relax}

\newcommand{\feii}{Fe$\;${\small\rm II}\relax}

\newcommand{\hit}{H$\;${\scriptsize\rm I}\relax}

\newcommand{\avg}[1]{\left< #1 \right>} 

\title[HVC/IVC covering factors and $z$-heights]{Intermediate- and high-velocity clouds in the Milky Way I: covering factors and vertical heights}

\author[Lehner et al.]{Nicolas Lehner$^1$, J. Christopher Howk$^1$, Antonino Marasco$^2$, Filippo Fraternali$^3$\\
$1$ Department of Physics and Astronomy, University of Notre Dame, Notre Dame, IN 46556, USA \\
$2$ INAF - Osservatorio Astrofisico di Arcetri, Largo E. Fermi 5, 50127, Firenze, Italy \\
$3$ Kapteyn Astronomical Institute, University of Groningen, Postbus 800, 9700 AV Groningen, The Netherlands \\
}

\date{Last updated \today; in original form \today}

\pubyear{2022}

\begin{document}
\label{firstpage}
\maketitle

\begin{abstract}
Intermediate- and high-velocity clouds (IVCs, HVCs) are a potential source of fuel for star formation in the Milky Way (MW), but their origins and fates depend sensitively on their distances. We search for IVC and HVC  in \hst\ high-resolution ultraviolet spectra of 55 halo stars at vertical heights $|z|\ga 1$ kpc. We show that  IVCs ($40 \le |\vlsr | <90$ \km) have a high detection rate---the covering factor, $f_c$---that is about constant ($f_c =0.90\pm 0.04$) from $z=1.5$ to 14 kpc, implying IVCs are essentially confined to $|z| \la 1.5$ kpc. For the HVCs  ($90 \le |\vlsr | \la 170$ \km), we find $f_c$ increases from $f_c \simeq 0.14\pm 0.10$ at  $|z|\la 2$--3 kpc to $f_c =0.60\pm 0.15$ at $5 \la |z|\la 14$ kpc, the latter value being similar to that found towards QSOs. In contrast, the covering factor of very high-velocity clouds (VHVCs, $|\vlsr | \ga 170$ \km) is $f_c<4\%$ in the stellar sample compared to 20\% in a QSO sample, implying these clouds must be at $d\ga 10$--15 kpc ($|z|\ga 10$ kpc). Gas clouds with $|\vlsr|>40$ \km\ at $|b|\ga 15\degr$ have therefore $|\vlsr|$  decreasing with decreasing $|z|$. Assuming each feature originates from a single cloud, we derive scale-heights of $1.0 \pm 0.3$ and $2.8 \pm 0.3$ kpc for the IVCs and HVCs, respectively. Our findings provide support to the ``rain" and galactic fountain models. In the latter scenario, VHVCs may mostly serve as fuel for the MW halo. In view of their locations and high covering factors, IVCs and HVCs are good candidates to sustain star formation in the MW.  
\end{abstract}
\begin{keywords}
Galaxy: halo -- Galaxy: kinematics and dynamics -- Galaxy: evolution -- Galaxy: structure -- ultraviolet: ISM
\end{keywords}


\section{Introduction}\label{s-intro}
The ability of the Milky Way (MW), like all star-forming galaxies, to form stars depends sensitively on the content and physical conditions of its gas. This in turn is dictated by internal effects, such as feedback, as well as by the external interactions of the Galaxy with its surroundings. The MW may gain mass from its circumgalactic environment through the infall of intergalactic matter or of remnants resulting from galaxy interactions, and it may lose mass through outflows driven by stellar feedback. These flows are critical for the MW's evolution, and they must result in a net gain of mass for the Galaxy to form stars over many several billion years (e.g., \citealt{chiappini01,schoenrich17}). An observational manifestation of such gas flows through the MW halo are likely the high-velocity clouds (HVCs), practically defined as gas with  $|\vlsr|\ge 90$ \km\ \citep[e.g.,][]{wakker97}.

Because the distances of the HVCs were unknown for a long time, the nature of these clouds has been debated for decades, with their origins variously attributed to Galactic fountain flows \citep[e.g.,][]{shapiro76,bregman80,fraternali06}, stripped or ejected gas from dwarf galaxies \citep[e.g.,][]{putman06,olano08}, condensations from the hot Galactic corona \citep[e.g.,][]{maller04,peek08,joung12,fraternali15}, gas in dark matter minihalos in the Local Group \citep[e.g.,][]{blitz99,giovanelli10}, or infalling intergalactic matter \citep{keres09}. The distance, $d$, to the HVCs is a critical parameter for distinguishing between streams of gas near the disc of the MW or Local Group phenomena, as well as for quantifying their basic physical properties, several of which directly scale with the distance (e.g., the mass $M \propto d^2$, see, e.g., \citealt{wakker97} for more details).

\begin{figure*}
\includegraphics[width=16 truecm]{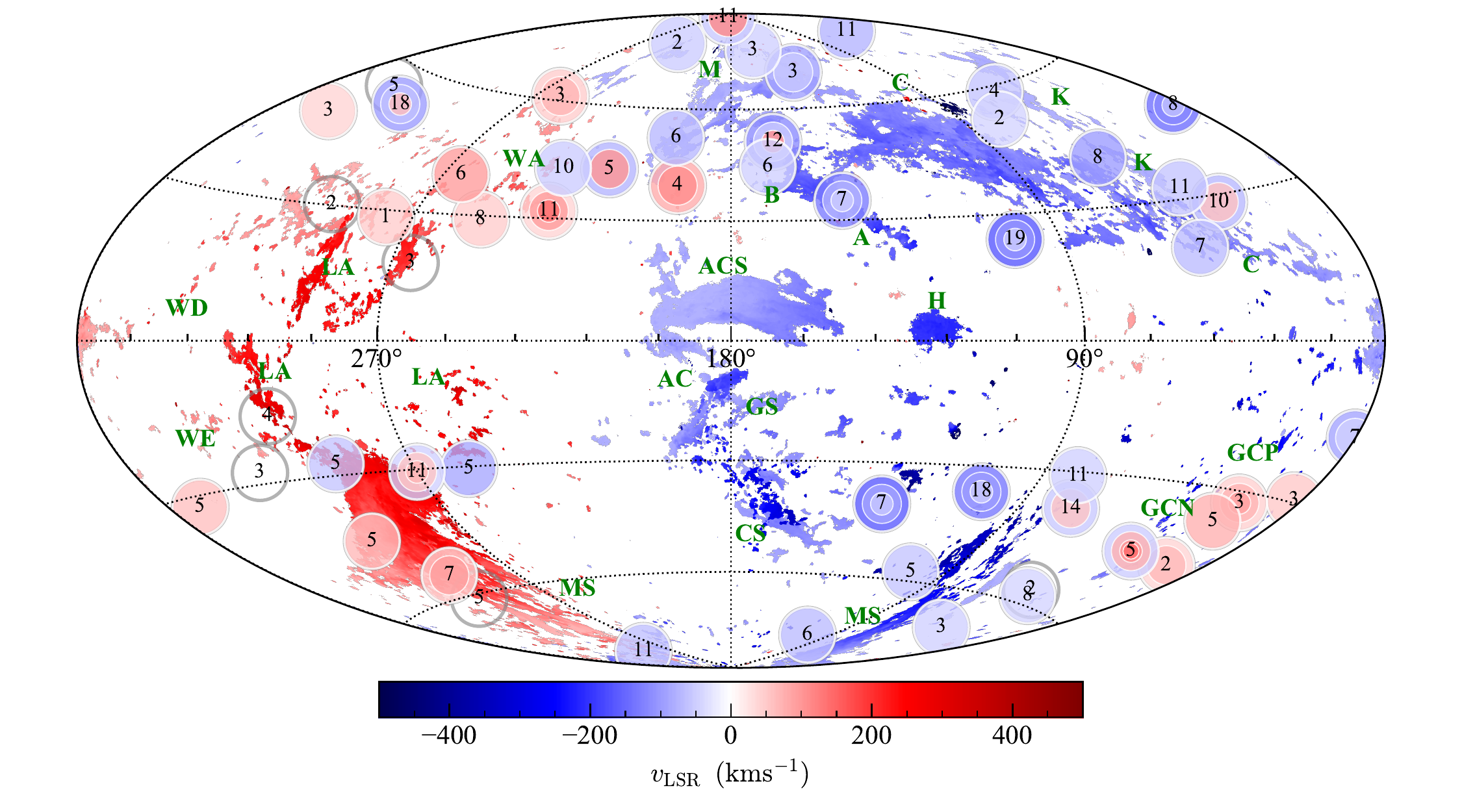}
\caption{Aitoff projection map of the Galactic HVCs from the 21-cm HI4PI survey \citep{westmeier18} with the distribution of our target stars on the Galactic sky (the IVC sky is not shown in this map). The \hit\ IVC sky is not shown on this map. Stellar sightlines with more than 1 HVC or IVC have multiple circles. The numbers inside the circles indicate the distances of the stars in kpc. Note that HVCs observed in front of the stars have only $90 \le |\vlsr|\la 170$ \km. }
\label{f-map-d}
\end{figure*}

Major efforts over the years have been undertaken to determine of the distances of specific \hi\ HVC complexes \citep[e.g.,][]{ryans97,wakker01,thom06,thom08,wakker07,wakker08,lehner10,smoker11}, which  are easily identified in the Galactic sky. About $13\%$--$18\%$ of the MW sky is covered with \hi\ HVCs at a sensitivity of $\mlnhi \ga 18.3$ (\citealt{wakker91,westmeier18}; see Fig~\ref{f-map-d}); a factor 2--3 deeper \hi\ emission observations yield a  covering factor of about 37\% \citep{murphy95,lockman02}. Although first thought predominantly neutral gas, H$\alpha$ emission and ultraviolet (UV) absorption observations have revealed an important ionized component for some of the \hi\ HVCs and intermediate-velocity clouds (IVCs, $40\la |\vlsr| \la 90$ \km) as well as in regions with previously no evidence of \hi\ HVCs or IVCs \citep[e.g.,][]{tufte98,haffner01,barger15,sembach99,sembach00,savage00,lehner01}. Surveys to search HVC absorption in QSOs spectra using  UV spectrographs, such as the {\it Far Ultraviolet Spectroscopic Explorer (FUSE)}\ and those onboard the {\it Hubble Space Telescope (HST)}, have in fact unveiled a very large covering fraction for the HVCs  \citep[e.g.,][]{sembach03,wakker03,fox06,shull09,richter09,lehner12,richter17}, demonstrating HVCs (and IVCs) are largely ionized clouds.  A recent HVC absorption spectroscopic survey with the Cosmic Origins Spectrograph (COS) on \hst\ used 270 QSOs to determine that the overall sky-covering fraction of high-velocity absorption using \siiii\ $\lambda$1206 is $f_c = 0.77 \pm 0.06$ \citep{richter17}, a factor 4 higher than \hi\ surveys. Many of these HVCs observed in absorption have been referred to as ionised HVCs (iHVCs, see \citealt{lehner11}, hereafter \citetalias{lehner11}) or highly ionised HVCs since in many cases a large fraction of the gas is photoionised or collisionally ionised, i.e.,  $\mnhii \gg \mnhi$. However, as discussed in \citet{lehner12} (hereafter \citetalias{lehner12}), some of these HVCs detected in  UV absorption can also be predominantly neutral. Therein we follow the \citetalias{lehner12} nomenclature: ``HVCs'' are clouds detected in absorption in the UV spectra of a background targets; ``\hi\ HVCs'' are clouds detected in 21-cm emission.

While high-resolution UV absorption spectroscopic surveys of the MW  have revealed that most of the Galactic sky is covered with HVCs, the diffuse ionised HVCs have not been mapped out, requiring  a different approach to determine their distances.  In a series of papers (\citealt{lehner10}; \citetalias{lehner11,lehner12}), Lehner et al. have studied  HVCs in absorption in the UV spectra of stars. The key advantage of using stars over QSOs is that the distance of the background targets is known, and hence an upper limit on distance of the clouds can be directly set. Noting the covering factor of HVCs towards QSOs is large, \citetalias{lehner11} hypothesised if HVCs are mostly within heliocentric distances of $d\sim 10$--20 kpc, they should have a similarly high covering fraction towards distant Galactic halo stars; if the HVCs were more distant, their covering factor should be much smaller. 

\citetalias{lehner11} (and see also \citetalias{lehner12}) observed 28 stars  at $|b|\ga 20\degr$ and vertical height $|z|\ga 3 $ kpc from the Galactic plane and heliocentric distances $3 \le d \la 32$ kpc (on average the stars are at $\langle d \rangle = 11.6 \pm 6.9$ kpc). The stars were chosen to be distributed across the high latitude Galactic sky, without {\it a priori}\ knowledge or checking if there was some known HVCs in these directions, i.e., avoiding  biases that could favor the presence or lack of HVCs along each sightline. \citetalias{lehner11} and \citetalias{lehner12} found that covering factors of the  HVCs with $90 \le |v_{\rm LSR}|\la 170$ \km\  in their QSO and stellar samples are not statistically distinguishable, implying that most of the HVCs with $90 \le |v_{\rm LSR}|\la 170$ \km\  are within $<5$--20 kpc of the sun.  These HVCs must therefore represent gas flows in the inner MW halo and could contain enough mass to potentially balance the MW's ongoing consumption (\citealt{shull09}; \citetalias{lehner11}; \citealt{marasco13}). On the other hand, no absorption from the VHVCs ($|\vlsr| \ga 170$ \km) was observed in the \citetalias{lehner11} stellar sample, suggesting this population lies at larger distances.

With their initial \hst\ survey, \citetalias{lehner11} did not have stars at $|z|<3$ kpc (by design), and therefore could not determine if the covering factor of HVCs may be significantly lower at vertical heights $|z|<3$ kpc, i.e., they could not determine if the HVCs may also be close to the MW disc or instead are largely located in a layer between the disc and halo, the so-called disc-halo interface. To address these questions, we present here a new \hst\ UV spectroscopic survey, using our and archival spectra of stars with accurate distances where we assembled a sample of 55 stars so that they are at $d\ga 2$ kpc and $1\la |z|\la 13$ kpc. The stars were chosen to have a good distribution on the Galactic sky irrespective of the location of known HVCs or IVCs either seen in \hi\ emission or absorption in QSO spectra. With this new sample, our goals are to determine the vertical distribution of the UV-absorbing gas in the Galactic halo, which we can infer by studying how the covering factors of HVCs and IVCs depend on $z$-height. These new constraints will be used in our second paper of the series (\citealt{marasco22}, hereafter \citetalias{marasco22}) to infer the 3D kinematics of the absorbing gas in order to directly derive the origins of the HVCs and IVCs.

Our paper is organised as follows. In \S\ref{s-data}, we present our sample of target stars and discuss the search and identification of the IVCs and HVCs from the analysis of the UV spectra. In \S\ref{s-main-res}, we present the main observational results from our analysis including the distances/$z$-heights of the IVCs and HVCs and the first study of the covering factors of the IVCs and HVCs as a function of $z$. In \S\ref{s-implications}, we discuss the VHVC population in the context of our findings and some implications, and finally in \S\ref{s-sum}, we summarise our main results.

\section{Sample, data, and basic analysis}\label{s-data}
\subsection{Sample description}\label{s-sample}

The bulk (46/55) of the stars in our sample comes from our HST cycle 17 and 20 programs (20 stars from PID 11592, 26 from PID 12982, respectively)\footnote{4 and 3 stars from our HST cycle 17 and 20 programs were removed because the S/N was too low.}; an additional 9 were found from previous  Space Telescope Imaging Spectrograph (STIS) observations in the \hst\ archive (see program identification, PID, in the last column of Table~\ref{t-sample}). Our stellar sample consists mostly of B-type stars (B1 to B5), post asymptotic giant branch (PAGB) stars, and blue horizontal branch (BHB) stars. \citetalias{lehner11} have demonstrated previously that these stars can be used to search effectively for interstellar HVC absorption using COS G130M/G160M and STIS E140M (see also, e.g., \citealt{howk03}; \citetalias{lehner12}; \citealt{werk19}).

The criteria for assembling the \citetalias{lehner11}'s stellar sample were that the stars are at $|z|\ga 3 $ kpc and are bright enough to be observed at a rate of $1$ \hst\ orbit per star. They favored broad spatial coverage (without regard to the \hi\ HVC sky locations) over a high density of stars in small regions of the sky. This contrasts for example from the survey of \citet{werk19}, where they target stars solely towards the north Galactic pole. For our Cycle 20 \hst\ program,  we adopted the same orbit requirement and similarly favored broad spatial coverage, but we selected stars with a lower vertical-height threshold $|z|\ge 1$ kpc in order to increase the sample of distant ($d> 2$ kpc) low-$z$ stars. 

\begin{figure}
\includegraphics[width=\columnwidth]{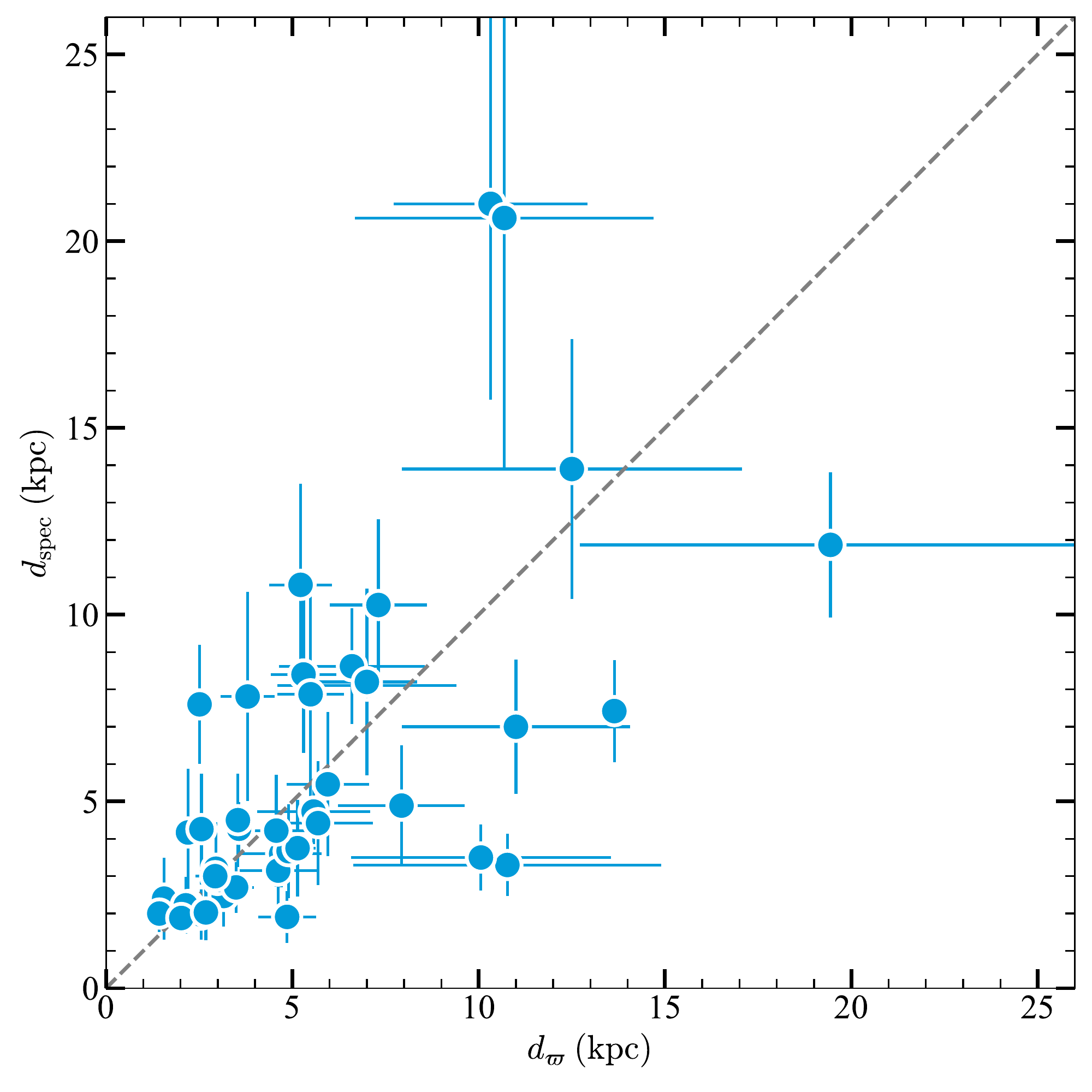}
\caption{Comparison of the stellar distances obtained from the stellar parameters ($y$-axis) and from the Gaia EDR3 parallaxes ($x$-axis) for the stars in our sample.}
\label{f-dist-comp}
\end{figure}

Our sample of stars is summarised in Table~\ref{t-sample}, which provides basic properties, including their Galactic coordinates, line-of-sight (radial) velocities and projected rotational velocities, spectroscopic distances and distances from parallax, and projected distances above or below the Galactic plane. All the references for the spectroscopic stellar distances are provided in Table~\ref{t-sample} and the reader should refer to these papers for a detailed description of this method. Stars associated with globular clusters (GCs) based on their location and velocity were placed at distances of the host GC. The method of \citet{chaboyer17} was used to estimate the distances of the GCs. In that method, main-sequence metal-poor stars with accurate parallaxes and abundance determinations are used to fit the main-sequence of the GCs to estimate the distances of the GCs (see \citealt{omalley17} for more detail). With that method, GC distances have typically accuracy better than $\la 12\%$. These methods to estimate the distances were used to assemble our initial sample, as very few stars had reliable parallaxes at that time. With the Gaia mission launched and the first data releases \citep{gaia16,gaia18,gaiadr3}, and especially Gaia early data release 3 (EDR3), the reliability of the parallaxes as distance estimators has, however, dramatically improved. There are now much more reliable parallaxes obtained for our entire sample, with the exception of the stars associated with GCs and PG0955+291.\footnote{For PG0955+291, the current Gaia EDR3 parallax remains completely unreliable with $0.036 \pm 0.041$ mas.} The advantage of using parallaxes is that the derived distances do not rely on specific stellar models (and possible systematic variations between different analyses). For the stars with Gaia parallaxes, we have on average $\langle \sigma_{d_{\varpi}}/d_{\varpi}\rangle = 0.22 \pm 0.15 $; however, 6/49 stars with parallaxes have $0.4\le  \sigma_{d_{\varpi}}/d_{\varpi} \le 0.6$. In Fig.~\ref{f-dist-comp}, we compare the stellar-model derived distances ($d_{\rm spec}$) and Gaia EDR3 distances ($d_{\varpi}$). In most cases, the stellar-model derived distances agree within 1--$2\sigma$ of the Gaia EDR3 distances, and therefore  results from our previous surveys (\citetalias{lehner11}; \citetalias{lehner12}) should hold (as we show here).  

\begin{figure*}
\includegraphics[width=16 truecm]{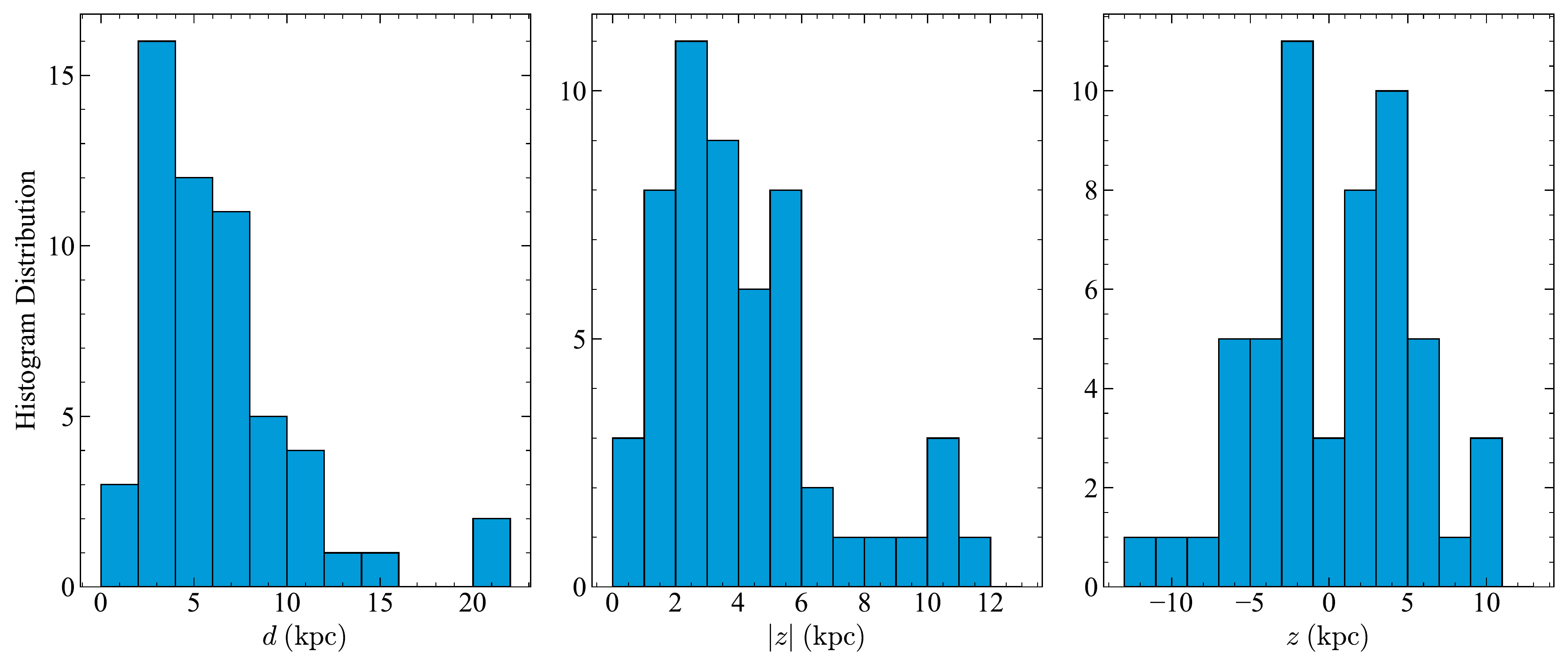}
\caption{Distribution of the distances ($d$) and  vertical $z$-heights of the stars in our sample.}
\label{f-dist}
\end{figure*}

For the distances and vertical heights ($z = d \sin b$), we therefore adopt the Gaia EDR3 distances except for stars associated with GCs and PG0955+291. In Fig.~\ref{f-dist}, we show the distribution of the stellar distances and vertical $z$-heights. Most (93\%) of the stars are within $d\la 12$ kpc. There is a good sampling of the vertical height up $|z| \simeq 7$ kpc as well as in both hemispheres up to that $z$-height. 

\begin{table*}
    \caption{Stellar sample \label{t-sample}}
    \scriptsize
    \centering
 \begin{minipage}{17.4 truecm}
    \begin{tabular}{lrrrrrrrlrr}
    \hline
\hline
\multicolumn{1}{c}{Name}& 
\multicolumn{1}{c}{$l$} &
\multicolumn{1}{c}{$b$}& 
\multicolumn{1}{c}{$v^{\star}_{\rm LSR}$}&   
\multicolumn{1}{c}{$v\sin i$}& 
\multicolumn{1}{c}{$d_{\rm spec}$}&  
\multicolumn{1}{c}{$d_{\varpi}$} &  
\multicolumn{1}{c}{$z$}& 
\multicolumn{1}{c}{Spectrograph} &
\multicolumn{1}{c}{PID} &
\multicolumn{1}{c}{Ref.} \\
\multicolumn{1}{c}{}& 
\multicolumn{1}{c}{($^\circ$)} &
\multicolumn{1}{c}{($^\circ$)} &
\multicolumn{1}{c}{(\km)}&   
\multicolumn{1}{c}{(\km)}&   
\multicolumn{1}{c}{(kpc)}&  
\multicolumn{1}{c}{(kpc)}&  
\multicolumn{1}{c}{(kpc)}&  
\multicolumn{1}{c}{} &  
\multicolumn{1}{c}{} &  
\multicolumn{1}{c}{} \\  
\hline
    NGC6723-III60  &	0.02 & $ -17.30 $ & $  -74.0 $ & $   0.0 $ & $  7.5 \pm   0.9 $ &      \nodata       & $  -2.2 \pm   0.3 $ &  COS G130M-G160M		       & 11592       &        15 \\
     EC19586-3823  &	2.33 & $ -29.61 $ & $  -96.0 $ & $ 150.0 $ & $  3.2 \pm   1.2 $ & $  3.0 \pm   0.4 $ & $  -1.5 \pm   0.2 $ & STIS E140M 	$ 0.2\times0.06$ & 12982       &	13 \\
          M5-ZNG1  &	3.86 & $  46.79 $ & $ + 65.7 $ & $   0.0 $ & $  7.9 \pm   0.7 $ &    \nodata	     & $   5.8 \pm   0.5 $ & STIS E140M 	$  0.2\times0.2$ &  9410       &	15 \\
         HD204076  &   13.90 & $ -45.68 $ & $  -13.0 $ & $ 102.0 $ & $  4.2 \pm   1.7 $ & $  2.2 \pm   0.3 $ & $  -1.6 \pm   0.2 $ & STIS E140H 	$ 0.1\times0.03$ & 12982       &	13 \\
         HD195455  &   20.27 & $ -32.14 $ & $ + 23.8 $ & $ 160.0 $ & $  2.0 \pm   0.7 $ & $  2.5 \pm   0.3 $ & $  -1.4 \pm   0.2 $ & STIS E140H 	$  0.2\times0.2$ &  8241       &	13 \\
     EC20485-2420  &   21.76 & $ -36.36 $ & $  -40.0 $ & $  10.0 $ & $  3.6 \pm   0.9 $ & $  4.7 \pm   1.1 $ & $  -2.8 \pm   0.6 $ & STIS E140M 	$  0.2\times0.2$ & 12982       &	10 \\
         HD206144  &   34.82 & $ -45.12 $ & $ +123.0 $ & $ 150.0 $ & $  3.2 \pm   1.2 $ & $  4.6 \pm   1.0 $ & $  -3.3 \pm   0.7 $ & STIS E140H 	$ 0.2\times0.09$ & 12982       &	13 \\
       PG1708+142  &   34.91 & $  28.47 $ & $  -28.0 $ & $  10.0 $ & $ 21.0 \pm   5.2 $ & $ 10.3 \pm   2.6 $ & $   4.9 \pm   1.2 $ &  COS G130M-G160M		       & 11592       &        1 \\
     EC23169-2235  &   40.89 & $ -68.63 $ & $ + 81.0 $ & $ 140.0 $ & $  2.5 \pm   0.8 $ & $  3.2 \pm   0.4 $ & $  -2.9 \pm   0.4 $ & STIS E140M 	$  0.2\times0.2$ & 12982       &	13 \\
           PHL346  &   41.17 & $ -58.15 $ & $ + 66.0 $ & $  45.0 $ & $  8.6 \pm   1.6 $ & $  7.9 \pm   3.3 $ & $  -6.7 \pm   2.8 $ & STIS E140M 	 $ 0.2\times0.2$ & 11592       &	13 \\
           VZ1128  &   42.50 & $  78.68 $ & $ -137.0 $ & $   0.0 $ & $ 10.7 \pm   0.9 $ &    \nodata	     & $  10.5 \pm   0.8 $ & STIS E140M 	 $  0.2\times0.06$ &  9150	 &	  15 \\
       PG1704+222  &   43.06 & $  32.36 $ & $  -22.0 $ & $  10.0 $ & $  7.0 \pm   1.8 $ & $ 11.0 \pm   3.1 $ & $   5.9 \pm   1.6 $ &  COS G130M-G160M		       & 11592       &        9 \\
         HD214080  &   44.80 & $ -56.92 $ & $ + 46.0 $ & $ 105.0 $ & $  2.4 \pm   1.1 $ & $  1.6 \pm   0.1 $ & $  -1.3 \pm   0.1 $ & STIS E140H 	 $0.1\times0.03$ & 12982       &	13 \\
         HD341617  &   50.67 & $  19.79 $ & $ + 63.0 $ & $  10.0 $ & $  8.1 \pm   2.0 $ & $  7.0 \pm   2.4 $ & $   2.4 \pm   0.8 $ &  COS G130M-G160M		       & 12982       &        9 \\
        Barnard29  &   58.97 & $  40.94 $ & $ -227.0 $ & $  20.0 $ & $  7.9 \pm   0.6 $ &	\nodata       & $   5.2 \pm   0.4 $ &  COS G130M-G160M		       & 11527       &       15 \\
       PG1511+367  &   59.21 & $  58.63 $ & $ +118.0 $ & $  77.0 $ & $  4.3 \pm   0.7 $ & $  3.6 \pm   0.4 $ & $   3.0 \pm   0.3 $ & STIS E140M 	 $ 0.2\times0.2$ & 11592       &	13 \\
       PG2219+094  &   73.16 & $ -38.40 $ & $  -17.0 $ & $ 225.0 $ & $  7.4 \pm   1.4 $ & $ 13.6 \pm   0.1 $ & $  -8.5 \pm   0.1 $ & STIS E140M 	 $ 0.2\times0.2$ & 11592       &	13 \\
       PG1533+467  &   75.18 & $  52.59 $ & $ + 50.0 $ & $ 215.0 $ & $  2.2 \pm   0.4 $ & $  2.2 \pm   0.2 $ & $   1.8 \pm   0.1 $ & STIS E140M 	 $ 0.2\times0.2$ & 12982       &	13 \\
       PG2214+184  &   79.40 & $ -30.90 $ & $  -70.0 $ & $  40.0 $ & $  3.3 \pm   0.8 $ & $ 10.8 \pm   4.1 $ & $  -5.5 \pm   2.1 $ &  COS G130M-G160M		       & 12982       &       11 \\
      HS1914+7139  &  102.99 & $  23.92 $ & $  -25.0 $ & $ 250.0 $ & $ 11.9 \pm   2.0 $ & $ 19.4 \pm   6.7 $ & $   7.9 \pm   2.7 $ &  COS G130M-G160M		       & 11592       &       13 \\
       PG0009+036  &  104.58 & $ -57.57 $ & $ +140.0 $ & $ 350.0 $ & $ 10.8 \pm   2.7 $ & $  5.2 \pm   0.8 $ & $  -4.4 \pm   0.7 $ &  COS G130M-G160M		       & 11592       &        6 \\
       PG2345+241  &  105.05 & $ -36.29 $ & $ + 85.0 $ & $  54.0 $ & $  4.9 \pm   1.6 $ & $ 18.0 \pm  10.8 $ & $ -10.7 \pm   6.4 $ & STIS E140M 	 $ 0.2\times0.2$ & 12982       &	13 \\
     EC00358-1516  &  108.17 & $ -77.49 $ & $ + 88.0 $ & $  35.0 $ & $  4.7 \pm   0.9 $ & $  5.6 \pm   1.5 $ & $  -5.4 \pm   1.5 $ & STIS E140M 	 $0.2\times0.06$ & 12982       &	13 \\
       PG0122+214  &  133.37 & $ -40.57 $ & $ + 24.0 $ & $ 117.0 $ & $  8.6 \pm   1.5 $ & $  6.6 \pm   2.0 $ & $  -4.3 \pm   1.3 $ & STIS E140M 	 $ 0.2\times0.2$ & 11592       &	13 \\
       PG1213+456  &  141.94 & $  70.44 $ & $  -15.0 $ & $  10.0 $ & $  2.9 \pm   0.7 $ & $  3.0 \pm   0.3 $ & $   2.8 \pm   0.3 $ &  COS G130M-G160M		       & 12982       &        2 \\
       PG0832+675  &  147.75 & $  35.01 $ & $  -67.0 $ & $   1.0 $ & $  8.2 \pm   2.5 $ & $  7.0 \pm   1.4 $ & $   4.0 \pm   0.8 $ &  COS G130M-G160M		       & 11592       &        3 \\
       PG1212+369  &  159.84 & $  77.72 $ & $  -32.0 $ & $ 100.0 $ & $  2.7 \pm   0.7 $ & $  3.5 \pm   0.5 $ & $   3.4 \pm   0.5 $ & STIS E140M 	 $0.2\times0.06$ & 12982       &	 2 \\
       PG1002+506  &  165.07 & $  50.94 $ & $ +  0.0 $ & $ 340.0 $ & $ 13.9 \pm   3.5 $ & $ 12.5 \pm   4.6 $ & $   9.7 \pm   3.5 $ &  COS G130M-G160M		       & 11592       &        5 \\
         HD233622  &  168.17 & $  44.23 $ & $ + 36.0 $ & $ 277.0 $ & $  4.4 \pm   1.7 $ & $  5.7 \pm   1.5 $ & $   4.0 \pm   1.0 $ & STIS E140M 	 $ 0.2\times0.2$ &  8662       &	13 \\
       PG0855+294  &  196.08 & $  39.12 $ & $ + 58.0 $ & $ 110.0 $ & $  7.8 \pm   2.8 $ & $  3.8 \pm   0.7 $ & $   2.4 \pm   0.5 $ & STIS E140M 	 $ 0.2\times0.2$ & 11592       &	13 \\
       PG0955+291  &  199.88 & $  51.94 $ & $ + 73.0 $ & $ 190.0 $ & $  6.1 \pm   1.9 $ &    \nodata	     & $   4.8 \pm   1.4 $ &  COS G130M-G160M		       & 11592       &       13 \\
       PG1243+275  &  206.51 & $  88.84 $ & $ +107.0 $ & $  10.0 $ & $  6.2 \pm   1.5 $ & $ 10.7 \pm   4.6 $ & $  10.7 \pm   4.6 $ &  COS G130M-G160M		       & 11592       &       10 \\
       PG0934+145  &  218.61 & $  43.08 $ & $ +105.0 $ & $  30.0 $ & $  8.4 \pm   2.1 $ & $  5.3 \pm   0.9 $ & $   3.6 \pm   0.6 $ &  COS G130M-G160M		       & 11592       &        6 \\
       PG0914+001  &  231.68 & $  31.84 $ & $ + 78.0 $ & $ 325.0 $ & $ 20.6 \pm   6.7 $ & $ 10.7 \pm   4.0 $ & $   5.6 \pm   2.1 $ &  COS G130M-G160M		       & 11592       &       13 \\
       PG0954+049  &  233.43 & $  42.79 $ & $ + 90.0 $ & $  10.0 $ & $  3.5 \pm   0.9 $ & $ 10.1 \pm   3.5 $ & $   6.8 \pm   2.4 $ &  COS G130M-G160M		       & 12982       &       10 \\
       PG1205+228  &  235.56 & $  79.12 $ & $ +157.0 $ & $ 165.0 $ & $  2.2 \pm   0.8 $ & $  2.1 \pm   0.2 $ & $   2.1 \pm   0.2 $ & STIS E140M 	 $0.2\times0.06$ & 12982       &	13 \\
     EC09452-1403  &  250.03 & $  29.16 $ & $ +224.0 $ & $  70.0 $ & $  4.9 \pm   1.6 $ & $  7.9 \pm   1.7 $ & $   3.9 \pm   0.8 $ &  COS G130M-G160M		       & 12982       &       13 \\
     EC05438-4741  &  254.39 & $ -30.53 $ & $ + 34.0 $ & $  30.0 $ & $  3.7 \pm   1.3 $ & $  4.9 \pm   0.5 $ & $  -2.5 \pm   0.2 $ &  COS G130M-G160M		       & 12982       &       13 \\
         HD100340  &  258.85 & $  61.23 $ & $ +255.2 $ & $ 156.0 $ & $  3.0 \pm   0.2 $ & $  2.9 \pm   0.5 $ & $   2.6 \pm   0.5 $ & STIS E140M 	 $ 0.2\times0.2$ &  8662       &	 7 \\
     EC10500-1358  &  264.36 & $  39.56 $ & $ + 91.0 $ & $ 100.0 $ & $  5.5 \pm   1.9 $ & $  5.9 \pm   1.1 $ & $   3.8 \pm   0.7 $ &  COS G130M-G160M		       & 11592       &       13 \\
          HD86248  &  264.59 & $  18.11 $ & $ + 60.0 $ & $  25.0 $ & $  7.6 \pm   1.6 $ & $  2.5 \pm   0.3 $ & $   0.8 \pm   0.1 $ & STIS E140H 	$ 0.1\times0.03$ & 12982       &	 8 \\
     EC05515-6107  &  270.08 & $ -30.61 $ & $ + 72.0 $ & $ 290.0 $ & $  4.0 \pm   1.3 $ & $ 10.8 \pm   4.7 $ & $  -5.5 \pm   2.4 $ & STIS E140M 	 $0.2\times0.06$ & 12982       &	13 \\
     EC11074-2912  &  277.82 & $  28.39 $ & $ + 45.0 $ & $  10.0 $ & $  2.0 \pm   0.5 $ & $  1.4 \pm   0.1 $ & $   0.7 \pm   0.0 $ &  COS G130M-G160M		       & 12982       &       10 \\
     EC06387-8045  &  292.58 & $ -27.43 $ & $ + 48.0 $ & $ 190.0 $ & $  3.8 \pm   1.3 $ & $  5.1 \pm   0.5 $ & $  -2.4 \pm   0.2 $ & STIS E140M 	 $ 0.2\times0.2$ & 12982       &	13 \\
         HD108230  &  296.76 & $  30.25 $ & $  -64.0 $ & $ 150.0 $ & \nodata & $  2.0 \pm   0.2 $ & $   1.0 \pm   0.1 $ & STIS E140H 	 $0.2\times0.09$ & 12982       &	16 \\
           LB3193  &  297.32 & $ -54.90 $ & $ + 98.0 $ & $  10.0 $ & $ 10.3 \pm   2.3 $ & $  7.3 \pm   1.3 $ & $  -6.0 \pm   1.1 $ & STIS E140M 	 $ 0.2\times0.2$ & 12982       &	14 \\
            SB357  &  300.94 & $ -80.78 $ & $ + 50.0 $ & $ 180.0 $ & $  6.0 \pm   1.0 $ & $ 11.3 \pm   5.0 $ & $ -11.2 \pm   5.0 $ &  COS G130M-G160M		       & 11592       &       13 \\
            JL212  &  303.63 & $ -61.03 $ & $ +  9.0 $ & $ 219.0 $ & $  1.9 \pm   0.7 $ & $  4.9 \pm   0.8 $ & $  -4.3 \pm   0.7 $ & STIS E140M 	 $ 0.2\times0.2$ &  8662       &	13 \\
         HD116852  &  304.88 & $ -16.13 $ & $  -55.0 $ & $ 136.0 $ & $  4.5 \pm   1.2 $ & $  3.5 \pm   0.4 $ & $  -1.0 \pm   0.1 $ & STIS E140H 	 $ 0.2\times0.2$ &  8241       &	12 \\
   NGC104-UIT14-2  &  305.93 & $ -44.87 $ & $  -29.0 $ & $   0.0 $ & $  4.9 \pm   0.6 $ &      \nodata       & $  -3.5 \pm   0.3 $ &  COS G130M-G160M		       & 11592       &        15 \\
       PG1323-086  &  317.11 & $  53.11 $ & $  -41.0 $ & $  10.0 $ & $ 15.8 \pm   3.7 $ & $ 17.7 \pm   8.9 $ & $  14.2 \pm   7.1 $ &  COS G130M-G160M		       & 11592       &        4 \\
     EC19071-7643  &  317.98 & $ -27.65 $ & $  -25.0 $ & $  30.0 $ & $  2.0 \pm   0.8 $ & $  2.7 \pm   0.2 $ & $  -1.2 \pm   0.1 $ & STIS E140H 	$ 0.2\times0.09$ & 12982       &	13 \\
         HD121968  &  333.97 & $  55.84 $ & $ + 49.0 $ & $ 200.0 $ & $  4.2 \pm   1.5 $ & $  4.6 \pm   1.1 $ & $   3.8 \pm   0.9 $ & STIS E140H 	$ 0.2\times0.09$ &  7270       &	13 \\
         HD125924  &  338.16 & $  48.28 $ & $ +249.0 $ & $  68.0 $ & $  4.3 \pm   1.5 $ & $  2.6 \pm   0.4 $ & $   1.9 \pm   0.3 $ & STIS E140H 	$ 0.2\times0.09$ & 12982       &	13 \\
     EC20011-5005  &  348.97 & $ -31.97 $ & $ -168.0 $ & $  30.0 $ & $  7.9 \pm   3.1 $ & $  5.5 \pm   0.9 $ & $  -2.9 \pm   0.5 $ &  COS G130M-G160M		       & 12982       &       13 \\
    \hline
\end{tabular}
    \footnotesize
Note: The stellar radial velocities of the stars are given in the LSR frame. The $d_{\rm spec}$ distances of stars are spectroscopic distances. The distances estimated by Gaia DR3, $d_{\varpi}$ are estimated using the parallax (\citealt{gaiadr3}). For this work we adopt for the distance, $d$, the parallax distances, except for stars in GCs that are based on the distance of the globular distance, and PG0955+291 for which the parallax is currently too uncertain to be used. The PID corresponds to the original \hst\ program identification number.
\\
References for stellar parameters and spectroscopic distances: (1) \citealt{conlon93}; (2) \citealt{ryans97}; (3) \citealt{ryans97a}; (4) \citealt{moehler98}; (5) \citealt{ringwald98}; (6) \citealt{rolleston99}; (7) \citealt{ryans99}; (8) \citealt{wakker01}; (9) 
\citealt{mooney02}; (10)  \citealt{smoker03}; (11) \citealt{lynn04}; (12) \citealt{bowen08}; (13) \citealt{silva11}; (14) \citealt{vickers15}; (15) \citealt{omalley17}; (16) \citealt{gaia16,gaia18,gaiadr3} (as well as the entire $d_{\rm Gaia}$ column).
\end{minipage}
\end{table*}
\normalsize

\subsection{\hst\ data}\label{s-hst-data}

For our HST programs, we used COS and STIS on \hst. All the archival data were drawn from the STIS echelle mode archive. COS G130M and G160M were used for 24 stars in our sample. This instrument setup provides a spectral resolution $R \sim 17,000$ (${\rm FWHM_{inst}} \simeq 18 $ \km). Thirty one stars were observed with the STIS echelle modes; the resolution of STIS varies between $R =45,800$ (6.5 \km\ with E140M) and $R=114,000$ (2.6 \km\ with E140H). Information about the COS instrument and COS spectra can be found in \citep{green12,massa13}. Information about STIS can be found in the STIS {\it HST}\ Instrument Handbook \citep{dressel07,branton21}. For both COS and STIS E140M, there is a near complete wavelength coverage between about 1150 and 1730 \AA. For STIS E140H, the wavelength coverage depends on the exact central wavelength setting, but for each star, there is coverage of the key ions in our study.  In these wavelength intervals, some or all of the following atomic and ionic species are available: \oi\ $\lambda$1302, \cii\ $\lambda$1334, \civ\ $\lambda\lambda$1548, 1550, \sii\ $\lambda\lambda$1250, 1253, 1259, \siii\ $\lambda$1190, 1193, 1260, 1304, 1526, \siiii\ $\lambda$1206, \siiv\ $\lambda\lambda$1393, 1402, \alii\ $\lambda$1670, \feii\ $\lambda$1608. All these atomic and ionic species were used to search high-velocity absorption components in the COS and STIS data.  

For the COS data, we retrieved the coadded data from the DR1 \hst\ Spectroscopic Legacy Archive (HSLA, \citealt{peeples17}).\footnote{Owing to the use of this new calibration, some of the velocities of the HVC observed in COS spectra reported in \citetalias{lehner11} and \citetalias{lehner12} can be somewhat different from those reported here.} For the STIS spectra, the proper alignment of the individual spectra was achieved through a cross-correlation technique. Data from individual exposures, including all grating configurations for echelle orders for STIS, were combined to produce a single spectrum with our in-house program. All the data were then shifted to the local standard of rest (LSR) frame. The absolute wavelength calibration of the STIS spectra is excellent, better than $\sim$1 \km, but that is not the case for COS, where large velocity shift is sometimes observed \citep[e.g.,][]{lehner15}. For the COS spectra,  we made a visual comparison between the Galactic \hi\ emission spectra from the Leiden/Argentine/Bonn (LAB) survey \citep{kalberla05} and the Galactic disc absorption of low ions (e.g., \sii) and atoms (\oi, \nni). In only one case (PG2214+184), we felt the need to shift the COS spectrum to better match the strong disc absorption with the \hi\ emission. 

The signal-to-noise (S/N) level is not uniform across the sample, but all the stellar spectra in our sample have S/N\,$> 10$ per resolution element. \citetalias{lehner12} showed that the detection of HVC absorption was not dependent on the S/N as long as S/N\,$> 3$ per resolution element due to the great strength of the transitions we use. Using the strong \cii\ $\lambda$1334 transition, we check how significant was the detection of \cii\ HVC or IVC absorption feature: for 95\% of the cases, the detections were $\gg 3\sigma$; only in 5\% of the cases, the detections were at the 3$\sigma$ level. The equivalent widths of \cii\ spreads between 20 and 252 m\AA, with 90\% being $>42$ m\AA, corresponding to the threshold of the robust sample in  \citetalias{lehner12}. The range of \cii\ equivalent widths in the stellar sample and QSO sample of \citetalias{lehner12} is quite similar as well (see their Fig. 4).

\subsection{Search for HVC and IVC absorption}\label{s-search}

In this paper, we are interested in determining the velocities of the IVCs and HVCs in order to compute their covering factors. The physical conditions of these clouds in the stellar sample will be addressed in the future. However, we note that the IVCs and HVCs must be largely ionised. For the HVCs in the \citetalias{lehner11} sample, the gas is ionised to levels $>50\%$--$95\%$ based on the [\oi/\siii] ratio. In order to search in the stellar spectra for IVC and HVC absorption features ($40 \le |v_{\rm LSR}|\le 400$ \km), we used strong transitions (e.g., \cii\ $\lambda$1334, \siii\ $\lambda$1260, and \siiii\ $\lambda$1206) and weaker transitions especially for the IVCs (e.g., \sii\ $\lambda\lambda$1250, \alii\ $\lambda$1670, \feii\ $\lambda$1608). Our rule to claim an IVC or a HVC detection is that absorption is observed in at least two atomic/ionic transitions of the same or different species. This follows the same definition as in \citetalias{lehner12} for the surveys of the HVC sky covering factor towards QSOs, but differs from \citet{richter17} where a single transition (e.g., \siiii\ $\lambda$1206 ) could be used to claim a HVC detection. 

\begin{figure}
\includegraphics[width=\columnwidth]{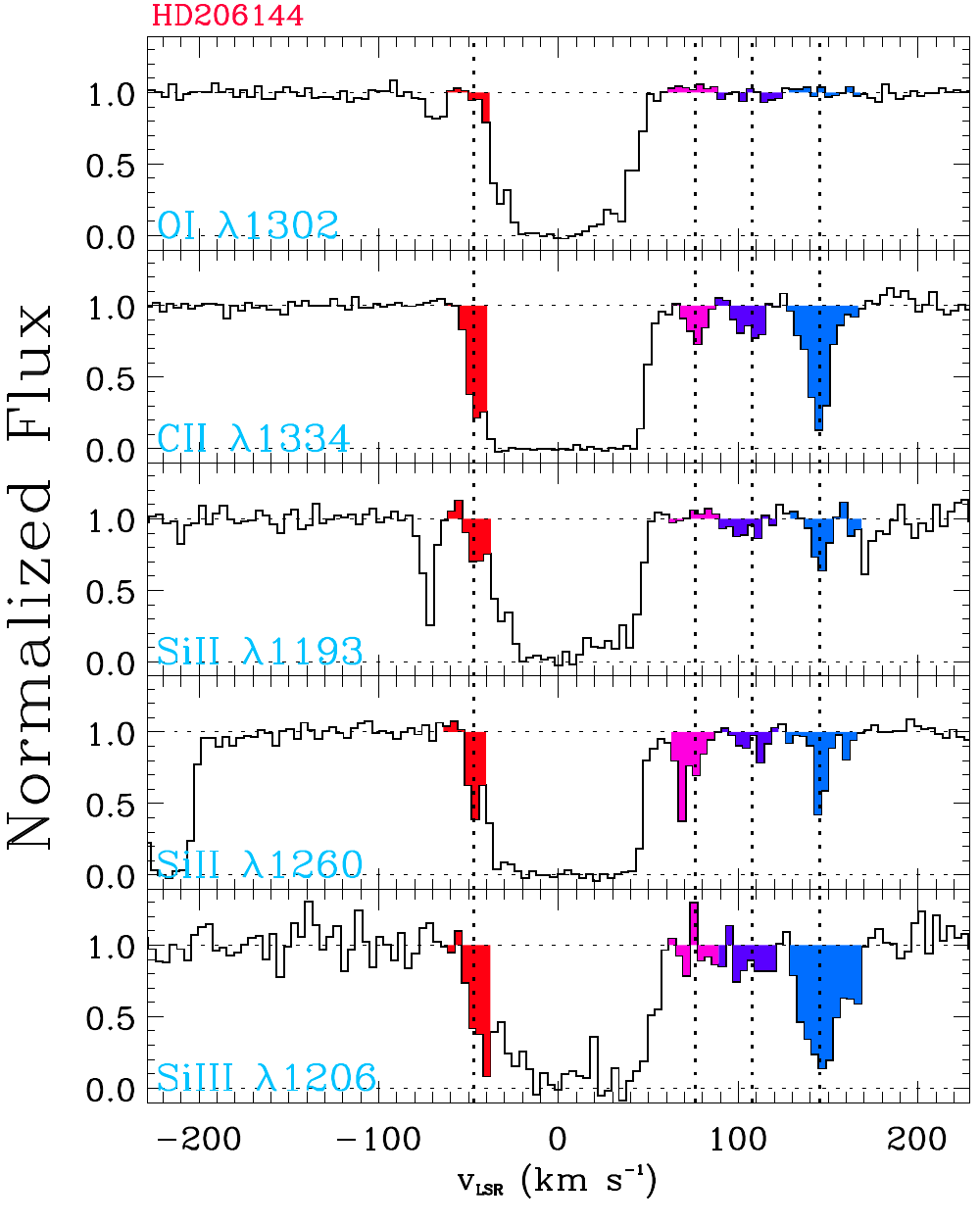}
\caption{Example of multiple IVC and HVC absorption features in one of the stars in our sample. The star HD206144 is at $d\simeq 4$ kpc and $z\simeq 3 $ kpc. The stellar velocity is at $\vlsr =+97$ \km, near one of the HVC components, but its $v\sin i = 64$ \km\ is large enough to ensure there is no contamination in the interstellar spectrum.   One can also observe additional absorption features from the MW disc at about $\vlsr \simeq -35,0,+30$ \km. }
\label{f-examp}
\end{figure}

In Fig.~\ref{f-examp}, we show an example of a STIS spectrum with the identifications of two IVC and two HVC absorption features. These are are highlighted in different colours, while the vertical dotted lines show their average central velocities. In this case, stellar contamination is not an issue because the projected rotational velocity, $v\sin i$, is large enough ($v \sin i = 64$ \km) to not confuse stellar and interstellar features (i.e., stellar features if present are much broader than the narrow interstellar lines, see below for more details). Several other examples of COS and STIS spectra can be found in \citet{lehner10}, \citetalias{lehner11}, and \citetalias{lehner12}. To determine the average velocities of the IVCs and HVCs in each species, we integrated the profiles over the observed absorption: $v = \int_{v_{1}}^{v_{2}} v \tau_a(v)dv /\int_{v_{1}}^{v_{2}} \tau_a(v)dv $, where $\tau_a(v)$ is the apparent optical depth profile. We then use a weighted average of the individual measurements to obtain our adopted velocity \vlsr\ of the IVCs or HVCs. All the identified IVCs and HVCs are reported in Table~\ref{t-result}, where we list the velocity integration ranges and average velocities of the IVCs and HVCs. 

As alluded to above, one concern with the use of stars as background continuum sources is the possible contamination from stellar photospheric lines. This is not issue when $v\sin i>30$ \km\ as the stellar lines can be clearly identified owing to their rotational broadening that is broader than the interstellar lines studied here. However, PAGB/BHB and evolved GC stars can have narrow photospheric lines that can mimic the narrow interstellar absorption features. For stars with $v\sin i<30$ \km, we therefore systematically check if contamination could be an issue.  There are 6 stars with $v\sin i<30$ \km\ in our sample where the interstellar and stellar velocities are within $\pm 20$ \km\ of each other. For one of them (LB3193), the spectrum was obtained with STIS, and the actual stellar absorption was previously observed at $+98 \pm 4$ \km\ \citep{mccausland92}, while our identification of a high-velocity absorption is at $+97$ \km, fully consistent with this stellar identification. The  $+97$ \km\ component has strong narrow, absorption of \oi, \siii, \cii, \siiii, and \siiv. The \siiv\ doublet is also saturated, which is common for stellar absorption but not in interstellar gas. Therefore, it is likely that the identified $+97$ \km\ component mostly arises from the star itself. The other 5 stars were observed with COS where the absolute velocity calibration is less accurate (see \S\ref{s-hst-data}):  NGC6723-III60 ($v^{\star}_{\rm LSR} = -74$ \km; $\vlsr = -88$ \km);  PG0832+675 ($v^{\star}_{\rm LSR} = -67$ \km; $\vlsr = -77,-50$ \km); PG1243+275 ($v^{\star}_{\rm LSR} = +107$ \km; $\vlsr = +114$ \km);  PG0934+145 ($v^{\star}_{\rm LSR} = +105$ \km; $\vlsr = +102$ \km); EC11074-2912 ($v^{\star}_{\rm LSR} = +45$ \km; $\vlsr = +49$ \km). NGC6723-III60, PG0934+145, and PG1243+275 are among the most metal rich stars in our sample and are discussed in \citetalias{lehner11}. For these stars, model atmospheres  were calculated by Dr. P. Chayer (priv. comm. 2011, see \citetalias{lehner11} for more details). For NGC6723-III60, the stellar model does not reproduce the observed HVC absorption, and therefore it is a genuine interstellar detection. For PG1243+275, the stellar model fills most of the absorption at $+114$ \km\ with stellar lines, but for \siii\ $\lambda$1526, the stellar spectrum is less crowded and both the stellar \siii\ at $+107$ \km\ and interstellar \siii\ at $+114$ \km\ are observed, giving us confidence in the detection of an HVC towards this star.  For PG0934+145, while the stellar model fills some of the absorption at $+102$ \km, there is room for interstellar absorption in  \oi, \sii\, \cii, or \alii, and therefore there is very likely a blend of both stellar and interstellar absorption in this spectrum too. The spectra of PG0832+675 is shown in Fig.~9 of \citetalias{lehner12}. The LAB survey survey shows there is a \hi\ emission detection at $-50$ \km, giving us confidence in the IVC detection at $-50$ \km. The other IVC at $-77$ \km\ is not observed in \hi; at the COS resolution and uncertainty in the COS wavelength calibration, it is plausible that this component could be associated with the star. Finally, for EC11074-2912, its $v\sin i$ is about 10 \km\ according to \citet{rolleston97}, and the close proximity in velocities strongly points to a stellar origin for the detections of \alii, \siii, \sii, and \feii\ at $+49$ \km. The $f^\star$ flag in the ninth column of Table~\ref{t-result} summarises the stellar contamination. Those marked with $f^\star = 2$ are not considered further in this work. 

We note that that IVC and HVC absorption lines can be found blueshifted {\it and} redshifted relative to the stars. Thus IVCs and HVCs are not circumstellar material (see, also, the ionization model and other arguments in \citealt{zech08}). The fact we can identify IVC/HVC absorption features in the most complicated stellar spectra gives us confidence that we have not failed to identify them when no intermediate- or high-velocity absorption is detected. That is there is truly no IVC or HVC absorption towards these stars. The following stars in our sample show no IVC and HVC absorption: HD214080, HD86248, HD108230, JL212, HD116852, EC19071-7643, and HD121968.

Finally, we note that the IVC absorption is less resolved and separated from the disc interstellar absorption in the COS spectra than in the STIS spectra owing to the lower spectral resolution of COS. However, this is not a detriment against detecting IVCs since we detect 26 IVCs in 24 COS spectra and 32 in 31 STIS spectra. For the HVCs, we find 13 and 10 in the COS and STIS spectra, respectively.

\begin{table*}
\caption{Velocities of detected IVCs and HVCs in absorption towards the stars \label{t-result}}
\tiny
\centering
\begin{minipage}{14.5 truecm}
\begin{tabular}{lrrrrrrrcl}
\hline
\multicolumn{1}{c}{Name}& 
\multicolumn{1}{c}{$l$} &
\multicolumn{1}{c}{$b$}& 
\multicolumn{1}{c}{$d$}&  
\multicolumn{1}{c}{$z$}& 
\multicolumn{1}{c}{ $v_1$}&   
\multicolumn{1}{c}{$v_2$}& 
\multicolumn{1}{c}{$\vlsr$} &
\multicolumn{1}{c}{$f_\star$} &
\multicolumn{1}{c}{Associated} \\
\multicolumn{1}{c}{}& 
\multicolumn{1}{c}{($^\circ$)} &
\multicolumn{1}{c}{($^\circ$)} &
\multicolumn{1}{c}{(kpc)}&  
\multicolumn{1}{c}{(kpc)}&  
\multicolumn{1}{c}{(\km)}&   
\multicolumn{1}{c}{(\km)}&   
\multicolumn{1}{c}{(\km)} &  
\multicolumn{1}{c}{} &  
\multicolumn{1}{c}{IVC/HVC complex} \\  
\hline
     NGC6723-III60 &    0.02 & $ -17.30 $ & $   7.5 $ & $ -2.2 $ & $-135.4 $ & $  -68.4 $ & $ -87.6 \pm    0.4 $ & 1 & None \\
     NGC6723-III60 &    0.02 & $ -17.30 $ & $   7.5 $ & $ -2.2 $ & $ -68.4 $ & $  -30.4 $ & $ -44.8 \pm    0.1 $ & 0 & None \\
      EC19586-3823 &    2.33 & $ -29.61 $ & $   3.0 $ & $ -1.5 $ & $  37.1 $ & $   92.1 $ & $  53.8 \pm    0.2 $ & 0 & None \\
           M5-ZNG1 &    3.86 & $  46.79 $ & $   7.9 $ & $  5.8 $ & $-162.0 $ & $ -126.0 $ & $-138.3 \pm    0.1 $ & 0 & Ionized complex L? \\
           M5-ZNG1 &    3.86 & $  46.79 $ & $   7.9 $ & $  5.8 $ & $-126.0 $ & $  -90.0 $ & $-114.4 \pm    0.1 $ & 0 & Ionized complex L? \\
           M5-ZNG1 &    3.86 & $  46.79 $ & $   7.9 $ & $  5.8 $ & $ -75.0 $ & $  -50.0 $ & $ -58.7 \pm    0.1 $ & 0 & Ionized complex L? \\
          HD204076 &   13.90 & $ -45.68 $ & $   2.2 $ & $ -1.6 $ & $  29.3 $ & $   54.3 $ & $  42.6 \pm    0.4 $ & 0 & None \\
          HD204076 &   13.90 & $ -45.68 $ & $   2.2 $ & $ -1.6 $ & $  54.3 $ & $   89.3 $ & $  70.2 \pm    0.2 $ & 0 & None \\
          HD195455 &   20.27 & $ -32.14 $ & $   2.5 $ & $ -1.4 $ & $  40.0 $ & $   52.0 $ & $  45.7 \pm    0.2 $ & 0 & None \\
          HD195455 &   20.27 & $ -32.14 $ & $   2.5 $ & $ -1.4 $ & $  61.0 $ & $   81.0 $ & $  71.7 \pm    0.1 $ & 0 & None \\
          HD195455 &   20.27 & $ -32.14 $ & $   2.5 $ & $ -1.4 $ & $  85.0 $ & $  102.0 $ & $  92.1 \pm    0.2 $ & 0 & None \\
      EC20485-2420 &   21.76 & $ -36.36 $ & $   4.7 $ & $ -2.8 $ & $  57.7 $ & $  102.7 $ & $  74.2 \pm    0.2 $ & 0 & None \\
          HD206144 &   34.82 & $ -45.12 $ & $   4.6 $ & $ -3.3 $ & $ -62.4 $ & $  -42.4 $ & $ -46.9 \pm    0.2 $ & 0 & Ionized PP Arch? \\
          HD206144 &   34.82 & $ -45.12 $ & $   4.6 $ & $ -3.3 $ & $  62.6 $ & $   87.6 $ & $  75.8 \pm    0.7 $ & 0 & Complex gp \\
          HD206144 &   34.82 & $ -45.12 $ & $   4.6 $ & $ -3.3 $ & $  87.6 $ & $  122.6 $ & $ 105.8 \pm    0.4 $ & 0 & Complex GCP\\
          HD206144 &   34.82 & $ -45.12 $ & $   4.6 $ & $ -3.3 $ & $ 127.6 $ & $  167.6 $ & $ 145.5 \pm    0.3 $ & 0 & None \\
        PG1708+142 &   34.91 & $  28.47 $ & $  10.3 $ & $  4.9 $ & $-127.2 $ & $  -47.2 $ & $ -67.8 \pm    0.5 $ & 0 & Complex K \\
        PG1708+142 &   34.91 & $  28.47 $ & $  10.3 $ & $  4.9 $ & $  42.8 $ & $   97.8 $ & $  61.6 \pm    0.4 $ & 0 & None \\
      EC23169-2235 &   40.89 & $ -68.63 $ & $   3.2 $ & $ -2.9 $ & $ -93.6 $ & $  -28.6 $ & $ -44.9 \pm    0.4 $ & 0 & Ionized PP Arch? \\
            PHL346 &   41.17 & $ -58.15 $ & $   7.9 $ & $ -6.7 $ & $ -57.7 $ & $  -32.7 $ & $ -41.9 \pm    0.1 $ & 0 & Ionized PP Arch? \\
            VZ1128 &   42.50 & $  78.68 $ & $  10.7 $ & $ 10.5 $ & $ -95.0 $ & $  -62.0 $ & $ -73.1 \pm    0.1 $ & 0 & Complex K? \\
        PG1704+222 &   43.06 & $  32.36 $ & $  11.0 $ & $  5.9 $ & $ -93.2 $ & $  -40.2 $ & $ -56.9 \pm    0.1 $ & 0 & Complex K \\
          HD341617 &   50.67 & $  19.79 $ & $   7.0 $ & $  2.4 $ & $-126.4 $ & $  -41.4 $ & $ -64.9 \pm    0.2 $ & 0 & Complex K \\
         Barnard29 &   58.97 & $  40.94 $ & $   7.9 $ & $  5.2 $ & $-140.0 $ & $  -75.0 $ & $ -97.6 \pm    0.1 $ & 0 & Complex C \\
        PG1511+367 &   59.21 & $  58.63 $ & $   3.6 $ & $  3.0 $ & $ -92.2 $ & $  -35.2 $ & $ -57.0 \pm    0.2 $ & 0 & Complex K \\
        PG2219+094 &   73.16 & $ -38.40 $ & $  13.6 $ & $ -8.5 $ & $ -96.8 $ & $  -36.8 $ & $ -50.9 \pm    0.3 $ & 0 & PP Arch\\
        PG2219+094 &   73.16 & $ -38.40 $ & $  13.6 $ & $ -8.5 $ & $  38.2 $ & $   83.2 $ & $  56.4 \pm    0.5 $ & 0 & Complex gp \\
        PG1533+467 &   75.18 & $  52.59 $ & $   2.2 $ & $  1.8 $ & $ -92.3 $ & $  -27.3 $ & $ -41.6 \pm    0.2 $ & 0 & Complex K/IV Arch \\
        PG2214+184 &   79.40 & $ -30.90 $ & $  10.8 $ & $ -5.5 $ & $ -76.0 $ & $  -33.0 $ & $ -45.4 \pm    0.6 $ & 0 & PP Arch \\
       HS1914+7139 &  102.99 & $  23.92 $ & $  19.4 $ & $  7.9 $ & $-190.8 $ & $ -145.8 $ & $-163.9 \pm    0.3 $ & 0 & Complex C \\
       HS1914+7139 &  102.99 & $  23.92 $ & $  19.4 $ & $  7.9 $ & $-140.8 $ & $  -75.8 $ & $-103.7 \pm    0.3 $ & 0 & Complex C/Outer Arm \\
       HS1914+7139 &  102.99 & $  23.92 $ & $  19.4 $ & $  7.9 $ & $ -79.3 $ & $  -32.7 $ & $ -48.1 \pm    0.5 $ & 0 & LLIV Arch/Outer Arm? \\
        PG0009+036 &  104.58 & $ -57.57 $ & $   5.2 $ & $ -4.4 $ & $ -91.3 $ & $  -41.3 $ & $ -56.3 \pm    0.2 $ & 0 & PP Arch \\
        PG2345+241 &  105.05 & $ -36.29 $ & $  18.0 $ & $-10.7 $ & $-151.0 $ & $ -111.0 $ & $-129.5 \pm    0.4 $ & 0 & None \\
        PG2345+241 &  105.05 & $ -36.29 $ & $  18.0 $ & $-10.7 $ & $-111.0 $ & $  -81.0 $ & $ -94.2 \pm    0.4 $ & 0 & None \\
        PG2345+241 &  105.05 & $ -36.29 $ & $  18.0 $ & $-10.7 $ & $ -67.0 $ & $  -36.0 $ & $ -47.3 \pm    0.4 $ & 0 & PP Arch \\
      EC00358-1516 &  108.17 & $ -77.49 $ & $   5.6 $ & $ -5.4 $ & $ -83.3 $ & $  -51.3 $ & $ -65.0 \pm    0.1 $ & 0 & PP Arch \\
        PG0122+214 &  133.37 & $ -40.57 $ & $   6.6 $ & $ -4.3 $ & $-184.0 $ & $ -134.0 $ & $-157.1 \pm    0.6 $ & 0 & WW503 \\
        PG0122+214 &  133.37 & $ -40.57 $ & $   6.6 $ & $ -4.3 $ & $-124.0 $ & $  -79.0 $ & $ -89.5 \pm    0.5 $ & 0 & Cohen Stream \\
        PG0122+214 &  133.37 & $ -40.57 $ & $   6.6 $ & $ -4.3 $ & $ -70.5 $ & $  -42.6 $ & $ -52.7 \pm    0.4 $ & 0 & PP Arch \\
        PG1213+456 &  141.94 & $  70.44 $ & $   3.0 $ & $  2.8 $ & $-120.4 $ & $  -80.4 $ & $ -92.1 \pm    0.2 $ & 0 & Complex M \\
        PG1213+456 &  141.94 & $  70.44 $ & $   3.0 $ & $  2.8 $ & $ -80.4 $ & $  -30.4 $ & $ -52.1 \pm    0.1 $ & 0 & IV Arch \\
        PG0832+675 &  147.75 & $  35.01 $ & $   7.0 $ & $  4.0 $ & $-159.1 $ & $ -104.1 $ & $-127.4 \pm    0.3 $ & 0 & Complex A \\
        PG0832+675 &  147.75 & $  35.01 $ & $   7.0 $ & $  4.0 $ & $-104.1 $ & $  -59.1 $ & $ -76.9 \pm    0.1 $ & 2 & IV Arch \\
        PG0832+675 &  147.75 & $  35.01 $ & $   7.0 $ & $  4.0 $ & $ -86.0 $ & $  -24.8 $ & $ -50.4 \pm    0.4 $ & 0 & IV Arch \\
        PG1212+369 &  159.84 & $  77.72 $ & $   3.5 $ & $  3.4 $ & $ -81.9 $ & $  -21.9 $ & $ -45.7 \pm    0.3 $ & 0 & IV Arch \\
        PG1002+506 &  165.07 & $  50.94 $ & $  12.5 $ & $  9.7 $ & $-152.3 $ & $  -92.3 $ & $-112.6 \pm    1.4 $ & 0 & Complex A/M \\
        PG1002+506 &  165.07 & $  50.94 $ & $  12.5 $ & $  9.7 $ & $ -82.3 $ & $  -32.3 $ & $ -49.7 \pm    0.8 $ & 0 & IV Arch \\
        PG1002+506 &  165.07 & $  50.94 $ & $  12.5 $ & $  9.7 $ & $  47.7 $ & $  197.7 $ & $ 112.1 \pm    2.0 $ & 0 & None \\
          HD233622 &  168.17 & $  44.23 $ & $   5.7 $ & $  4.0 $ & $ -86.8 $ & $  -21.8 $ & $ -46.8 \pm    0.1 $ & 0 & IV Arch \\
        PG0855+294 &  196.08 & $  39.12 $ & $   3.8 $ & $  2.4 $ & $  57.0 $ & $   94.0 $ & $  79.7 \pm    0.3 $ & 0 & None \\
        PG0855+294 &  196.08 & $  39.12 $ & $   3.8 $ & $  2.4 $ & $  94.0 $ & $  137.0 $ & $ 109.5 \pm    0.2 $ & 0 & Cloud WW113 \\
        PG0955+291 &  199.88 & $  51.94 $ & $   6.1 $ & $  4.8 $ & $-102.1 $ & $  -52.1 $ & $ -69.2 \pm    0.3 $ & 0 & IV Arch \\
        PG1243+275 &  206.51 & $  88.84 $ & $  10.7 $ & $ 10.7 $ & $ -82.0 $ & $  -37.0 $ & $ -51.2 \pm    0.1 $ & 0 & IV Spur \\
        PG1243+275 &  206.51 & $  88.84 $ & $  10.7 $ & $ 10.7 $ & $  83.0 $ & $  153.0 $ & $ 114.0 \pm    0.5 $ & 1 & None \\
        PG0934+145 &  218.61 & $  43.08 $ & $   5.3 $ & $  3.6 $ & $-108.1 $ & $  -58.1 $ & $ -72.9 \pm    0.3 $ & 0 & IV Arch/IV Spur \\
        PG0934+145 &  218.61 & $  43.08 $ & $   5.3 $ & $  3.6 $ & $  66.9 $ & $  141.9 $ & $ 102.0 \pm    0.2 $ & 1 & Complex WB \\
        PG0914+001 &  231.68 & $  31.84 $ & $  10.7 $ & $  5.6 $ & $  23.6 $ & $   73.6 $ & $  42.9 \pm    0.5 $ & 0 & None \\
        PG0914+001 &  231.68 & $  31.84 $ & $  10.7 $ & $  5.6 $ & $  73.6 $ & $  123.6 $ & $  97.1 \pm    0.4 $ & 0 & Complex WB \\
        PG0914+001 &  231.68 & $  31.84 $ & $  10.7 $ & $  5.6 $ & $ 123.6 $ & $  163.6 $ & $ 143.4 \pm    1.3 $ & 0 & Complex WA/WB \\
        PG0954+049 &  233.43 & $  42.79 $ & $  10.1 $ & $  6.8 $ & $ -73.7 $ & $  -33.7 $ & $ -49.4 \pm    0.2 $ & 0 & Ionized IV Spur? \\
        PG1205+228 &  235.56 & $  79.12 $ & $   2.1 $ & $  2.1 $ & $ -71.2 $ & $  -37.3 $ & $ -47.3 \pm    0.2 $ & 0 & IV Spur \\
      EC09452-1403 &  250.03 & $  29.16 $ & $   7.9 $ & $  3.9 $ & $  31.5 $ & $   75.5 $ & $  48.3 \pm    0.3 $ & 0 & Ionized IV-WA? \\
      EC05438-4741 &  254.39 & $ -30.53 $ & $   4.9 $ & $ -2.5 $ & $-126.9 $ & $  -61.9 $ & $ -83.6 \pm    0.5 $ & 0 & Nond \\
          HD100340 &  258.85 & $  61.23 $ & $   2.9 $ & $  2.6 $ & $  40.0 $ & $   60.0 $ & $  50.1 \pm    0.1 $ & 0 & IV-WA \\
          HD100340 &  258.85 & $  61.23 $ & $   2.9 $ & $  2.6 $ & $  60.0 $ & $   90.0 $ & $  72.0 \pm    0.1 $ & 0 & IV-WA \\
      EC10500-1358 &  264.36 & $  39.56 $ & $   5.9 $ & $  3.8 $ & $  66.3 $ & $  121.3 $ & $  94.5 \pm    3.8 $ & 0 & Cloud WW95 \\
      EC05515-6107 &  270.08 & $ -30.61 $ & $  10.8 $ & $ -5.5 $ & $ -76.3 $ & $  -26.3 $ & $ -46.2 \pm    0.1 $ & 0 & None \\
      EC05515-6107 &  270.08 & $ -30.61 $ & $  10.8 $ & $ -5.5 $ & $  43.7 $ & $   66.7 $ & $  55.9 \pm    0.1 $ & 0 & None \\
      EC05515-6107 &  270.08 & $ -30.61 $ & $  10.8 $ & $ -5.5 $ & $  66.7 $ & $   88.7 $ & $  73.9 \pm    0.1 $ & 0 & None \\
      EC11074-2912 &  277.82 & $  28.39 $ & $   1.4 $ & $  0.7 $ & $  33.2 $ & $   72.2 $ & $  48.9 \pm    0.5 $ & 2 & Ionized IV-WA? \\
      EC06387-8045 &  292.58 & $ -27.43 $ & $   5.1 $ & $ -2.4 $ & $ -86.8 $ & $  -46.8 $ & $ -57.1 \pm    0.2 $ & 0 & None \\
            LB3193 &  297.32 & $ -54.90 $ & $   7.3 $ & $ -6.0 $ & $  29.9 $ & $   59.9 $ & $  43.1 \pm    0.2 $ & 0 & None \\
            LB3193 &  297.32 & $ -54.90 $ & $   7.3 $ & $ -6.0 $ & $  74.9 $ & $  134.9 $ & $  96.7 \pm    0.3 $ & 2 & None \\
             SB357 &  300.94 & $ -80.78 $ & $  11.3 $ & $-11.2 $ & $ -86.9 $ & $  -46.9 $ & $ -58.6 \pm    0.4 $ & 0 & None \\
    NGC104-UIT14-2 &  305.93 & $ -44.87 $ & $   4.9 $ & $ -3.5 $ & $  50.0 $ & $  100.0 $ & $  67.4 \pm    0.6 $ & 0 & None \\
        PG1323-086 &  317.11 & $  53.11 $ & $  17.7 $ & $ 14.2 $ & $-140.0 $ & $  -75.0 $ & $ -94.8 \pm    0.4 $ & 0 & None \\
        PG1323-086 &  317.11 & $  53.11 $ & $  17.7 $ & $ 14.2 $ & $ -75.0 $ & $  -50.0 $ & $ -61.0 \pm    0.1 $ & 0 & None \\
        PG1323-086 &  317.11 & $  53.11 $ & $  17.7 $ & $ 14.2 $ & $  50.0 $ & $  100.0 $ & $  62.7 \pm    0.3 $ & 0 & IV-WA \\
          HD125924 &  338.16 & $  48.28 $ & $   2.6 $ & $  1.9 $ & $  33.0 $ & $   55.0 $ & $  40.4 \pm    0.1 $ & 0 & None \\
      EC20011-5005 &  348.97 & $ -31.97 $ & $   5.5 $ & $ -2.9 $ & $  45.0 $ & $  100.0 $ & $  62.1 \pm    0.3 $ & 0 & None \\
\hline
\end{tabular}
    \footnotesize
Note: The velocities $v_1$ and $v_2$ correspond to the integration range of each velocity component and \vlsr\ is the average velocity of each IVC or HVC. For each absorption feature, we checked potential contamination from stellar photosphere when $v\sin i \le 30$ \km\ and $v_\star -20 \le \vlsr \le v_\star +20 $ \km.  The flag $f_\star$ informs us on that plausible contamination: $f_\star = 0$ implies no stellar contamination; $f_\star = 1$ implies some stellar contamination, but some absorption is still due to the ISM; $f_\star = 2$ implies that most of the absorption is likely due to the star. When $f_\star = 2$, this detection is not considered further. In the last column, we indicate the potential \hit\ complexes towards the stars based on their velocity-position from \citet{wakker04a}; if the position is not quite in the range give in \citet{wakker04a} but the LSR velocities are similar, we mark these directions as potentially ionised regions of these complexes.    
\end{minipage}
\end{table*}
\normalsize

\section{Covering factors and distances of the IVCs and HVCs}\label{s-main-res}
\subsection{General comments}\label{s-comments}
In our sample of 55 stars at $|z| \ga 1$ kpc, we find 23 HVCs and 58 IVCs. In Fig.~\ref{f-map-d}, we show the Galactic sky distribution of the detections and non-detections overlaid with the \hi\ 21-cm emission map for the HVCs. The number in each circle indicates the distance of the star in kpc, which provides  a strict upper limit on the distance of the IVC or HVC when detected in absorption towards the given star. In the last column of Table~\ref{t-result}, we give the names of the main \hi\ HVC and IVC complexes  following the nomenclature from \citet{wakker04a} where we match the positions and LSR velocities with the \hi\ surveys. For IVCs or HVCs that are close to \hi\ complexes and have similar velocities, we mark those as potential ionised gas associated with the \hi\ complexes. About 48\% of the HVCs and 55\% of the IVCs detected towards the stars are identified with known \hi\ complexes. About 10\% in each category are likely ionised envelopes to these \hi\ complexes. The remaining 42\% of the HVCs and 35\% of the IVCs are not associated with catalogued \hi\ complexes and therefore must be largely ionised. 

To estimate the covering factor for the IVCs and HVCs, we  assume a binomial distribution. We follow \citet{cameron11} in assessing the likelihood function for values of the covering factor, given the number of detections (successes) against the total sample (number of stars at a given $z$-height) (see \citealt{howk17} for more detail). The confidence intervals quoted here are at the 68\% level. While a sightline may have more than one high-velocity absorption component, only one HVC or IVC for a given sightline is counted for estimating the covering factor of the HVC and IVC, respectively. This follows the methodology adopted in our previous works (\citetalias{lehner11,lehner12}), but not necessarily other estimates. For example, \citet{richter17} use a single ion to confirm a HVC detection, but also account of multiple HVCs to estimate $f_c$ if there is more than one along a given sightline. For our controlled sample of QSOs, we use the results from \citetalias{lehner12} owing to the easy access to the velocity information and the methodology to search for HVC detection that is the same as our present study. However, we note that for the HVCs+VHVCs, the larger survey from \citet{richter17} using the same methodology than ours to estimate $f_c$  gives a very similar covering factor: $f_c = 0.66 \pm 0.02$ using the information from \citet{richter17} compared to $f_c = 0.65 \pm 0.04$ in \citetalias{lehner12}.

\subsection{HVC covering factor as a function of $z$}\label{s-fc-hvc}
\begin{figure}
\includegraphics[width=\columnwidth]{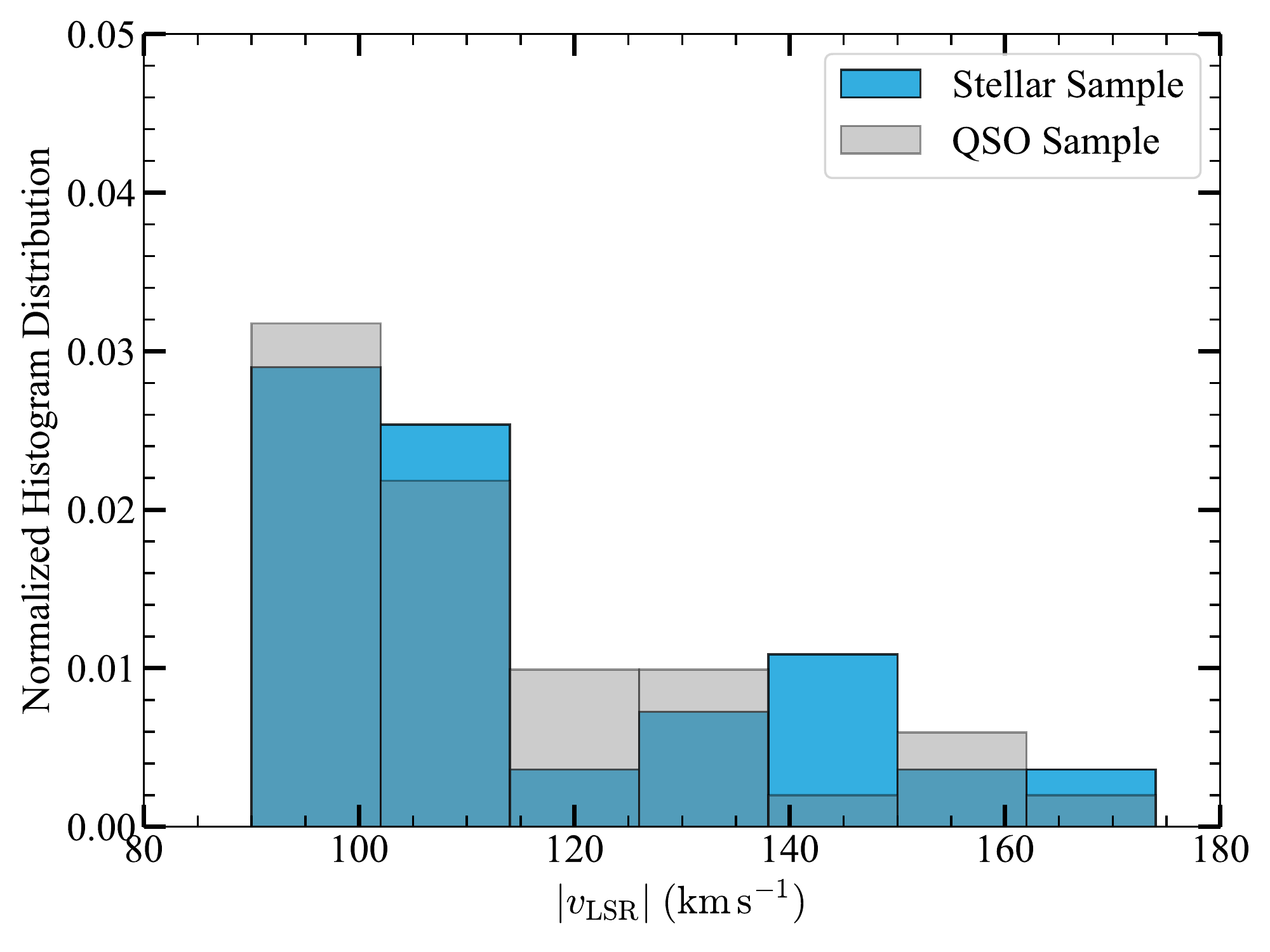}
\caption{Comparison of the absolute velocity distribution observed in the stellar (this paper) and QSO  (from \citetalias{lehner12}) samples for HVCs with $90 \le |\vlsr|\le 170$  \km.}
\label{f-hist-qso-star-vel}
\end{figure}

Prior to estimating the covering factor of the HVCs, we show in Fig.~\ref{f-hist-qso-star-vel} the distribution of the absolute LSR velocities of  HVCs in the stellar and QSO \citetalias{lehner12} samples, demonstrating an excellent overlap for the velocity distributions between the two samples. A two-sample Kolmogorov-Smirnov (KS) test statistically confirms the visual impression that we cannot rule out the null hypothesis that there is no difference between the two distributions. Since the QSOs and stars were selected to cover the Galactic sky irrespective of the distribution of the \hi\ 21-cm emission HVCs, it is remarkable that the velocity distributions in both samples are quite similar, and it is therefore appropriate to compare the covering factors between the two samples.

\citetalias{lehner11} and \citetalias{lehner12} have 28 stars  at $\avg{|z|} =6.5 \pm 3.2$ kpc in their sample, and found that the covering factors of the  HVCs  in the QSO and stellar samples were $f_c = 0.59 \pm 0.04$ and $f_c = 0.50 \pm 0.09$, respectively. These were not statistically different, but there was a hint that the covering of the HVCs in the stellar sample was somewhat smaller, likely owing to a large fraction of the non-detections being found in stars at $|z|\simeq 3 $ kpc, a vertical height where the covering factor of the HVCs may dip. Now, with a large sample of stars and probing low-$z$ stars, we can robustly assess that tentative conclusion. 

\begin{figure}
\includegraphics[width=\columnwidth]{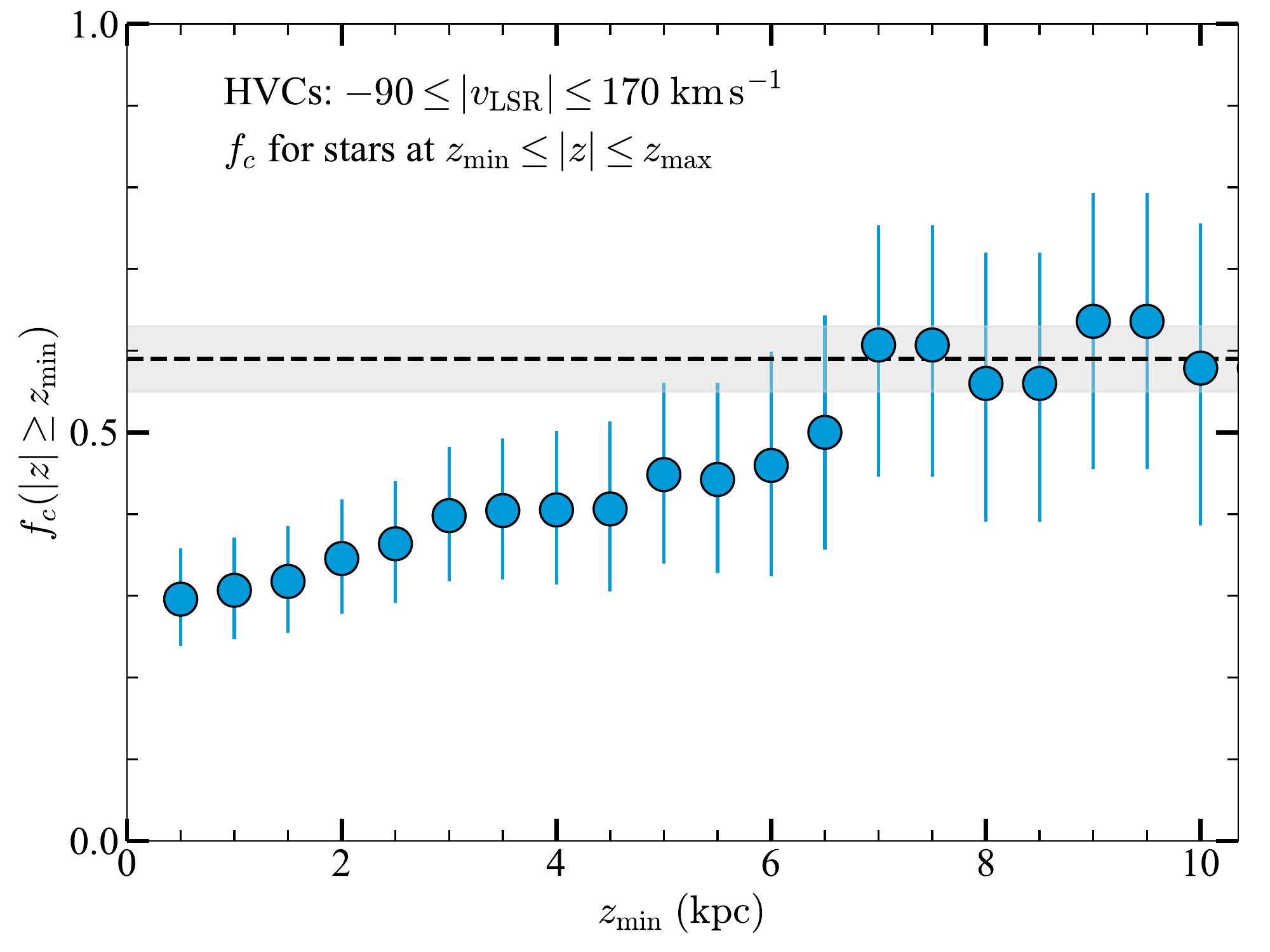}
\caption{Covering factors of the HVCs as a function of the $z$-height determined between $z_{\rm min}  =0.5,...,10$ kpc and $z_{\rm max}= 14.2$ kpc (the maximum $z$-height in our stellar sample). The dashed line (and 1$\sigma$ error bar shown with the gray area) shows the mean covering factor of the HVCs with $90 \le |\vlsr|\le 170$  \km\ determined from the QSO sample \citepalias{lehner12}. The confidence intervals are at the 68\% level.  }
\label{f-fc-hvc}
\end{figure}

In  Fig.~\ref{f-fc-hvc}, we  show the covering factors of the HVCs towards the stars, where we estimate $f_c$ in intervals of $z$-height between a given minimum $|z|$-height value and the  maximum $|z|$-height in our stellar sample. In that figure, we also show the covering factor of the HVCs towards QSOs from \citetalias{lehner12} with the horizontal dashed line and its $1\sigma$ error with the gray shaded area. Fig.~\ref{f-fc-hvc} shows a clear increase of $f_c$ with increasing $z_{\rm min}$, i.e., as more stars at low $z$-height are removed from the sample, the covering factor of the HVCs becomes larger. Around $z_{\rm min} \simeq 6.5$ kpc, the covering factor plateaus to a value consistent with the covering factor of HVCs seen towards QSOs. Within a $1\sigma$ confidence overlap, the covering factors towards QSOs and stars at $z_{\rm min} \simeq 5$ kpc are already consistent with each other. 

Another way to study the change of $f_c$ with $z$ is to specifically estimate the covering factors in several $z$ intervals. Using only stars at $1<|z|<3$ kpc, we find $f_c(|z|<3\,{\rm kpc}) = 0.15 \pm 0.08$ (3/22) and for $1<|z|< 2$ kpc, $f_c(|z|<2 \,{\rm kpc}) = 0.14\,^{+0.11}_{-0.08}$ (1/11). On the other hand, for stars with $|z|\ge 3$ kpc, we find $f_c(|z|\ge 3\,{\rm kpc}) = 0.40 \pm 0.08$ (13/33), for $|z|\ge 5$ kpc, $f_c(|z|\ge 5\,{\rm kpc}) = 0.45 \pm 0.11$ (8/18), and for  $|z|\ge 7$ kpc, $f_c(|z|\ge 7\,{\rm kpc}) = 0.60 \pm 0.15$ (5/8).\footnote{In \citetalias{lehner11}, we found a somewhat higher covering factor with $f_c = 0.50 \pm 0.09$ for $|z|\ga 3$ kpc stars. However, if we remove the uncertain sightlines as we did in the present work and use strictly $|z|\ge 3$ kpc, we would find $f_c = 0.43 \pm 0.10$, fully consistent with the new sample present here (and within $1\sigma$ from the estimate in \citetalias{lehner11}).} Therefore a similar decrease in the covering factor of HVCs as $z$ decreases is again observed. However, there is also still a non-negligible amount of HVCs at lower $z$ height: considering the stars at $3\le |z| \le 6$ kpc, we find $f_c = 0.36 \pm 0.10$, which is about 60\% of the covering factor observed towards QSOs. 

Hence considering  all these results, it is apparent that HVCs are not as frequently observed in stellar spectra of stars at $|z|<3$ kpc as in higher $z$ stars. At $3\la |z| \la 6$ kpc, HVCs become more common, but still not as widespread as observed towards the QSOs. Only when the stars have  $6\la |z| \la 14$ kpc, the covering factor of the HVCs in the stellar sample becomes fully consistent with that observed towards QSOs. 

Our sample of 16 sightlines with HVCs (counting  only 1 HVC per sightline to estimate $f_c$) is  small for testing differences  between the two Galactic hemispheres. Nevertheless, we note there are 12  HVC sightlines in the northern hemisphere and 4 in the southern hemisphere, while the sample has 30 stars at $b>0\degr$ and 25 stars at $b<0\degr$, i.e., about the same number of stars in both hemispheres. There is therefore a hint that the covering factor of the HVCs in the northern hemisphere is about twice larger than that in southern hemisphere. However, the values of $f_c$ overlap within about $1 \sigma$ in both hemispheres. Also complicating the comparison,  the northern hemisphere sample is more uniformly spread than the southern hemisphere sample (see Fig.~\ref{f-map-d}). If we consider all the stars at $b>0\degr$, we note that the results above hold, but the covering factors of the QSOs would be recovered at somewhat smaller $z$-height since $f_c =0.61 \pm 0.12$ for  $|z|\ga 4$ kpc.

\subsection{IVC Covering factor as a function of $z$}\label{s-fc-ivc}
Contrary to the HVCs, IVCs have not been studied as systematically, especially in the UV bandpass. Their covering factor from UV absorption line studies towards QSOs has not previously been calculated. Based on \hi\ emission observations, the IVC \hi\ covering factor is quite large with $f_c \simeq 0.4$ for a sensitivity level of $\mlnhi \ga 19$ (\citealt{richter06}), about 2.5 times larger than that of the HVCs (see \S\ref{s-intro}). An optical survey towards QSOs using \caii\ and \nai\ shows IVC detection rates of 40\%  and 25\%, respectively \citep{benbekhti12}. In 50\% of the cases where \caii\ is detected, they are associated with major structures in both Galactic hemispheres, such as the IV Arch, LLIV Arch, IV Spur, complex K in the northern hemisphere, PP Arch or IV South and complex gp in the southern hemisphere (see, e.g., \citealt{wakker04a}). These complexes are also known to be predominantly in the lower Galactic halo based on detections of \caii\ absorption towards stars: for example, the IV-Arch is at $z<1.7$ kpc, IV Spur at $z<2.1$ kpc, LLIV Arch at $z<1.9$ kpc, complex K at $z<4.5$ kpc, PP Arch at $z<0.9$ kpc, complex gp at $z<2$ kpc (\citealt{wakker01,wakker04a} and references therein). However, there is still a wide range of $z$ allowed with these limits, and only the PP Arch is clearly at a low vertical height with $z<0.9$ kpc. Also as discussed in \S\ref{s-comments} and shown in Table~\ref{t-result}, while about 55\% of the IVCs probed by our sample are associated with one of these complexes, the other 45\% are not catalogued \hi\ 21-cm IVCs. Therefore, our sample provides new constraints on the IVC distances and their covering factor especially since we use dominant ions to identify them. 

\begin{figure}
\includegraphics[width=\columnwidth]{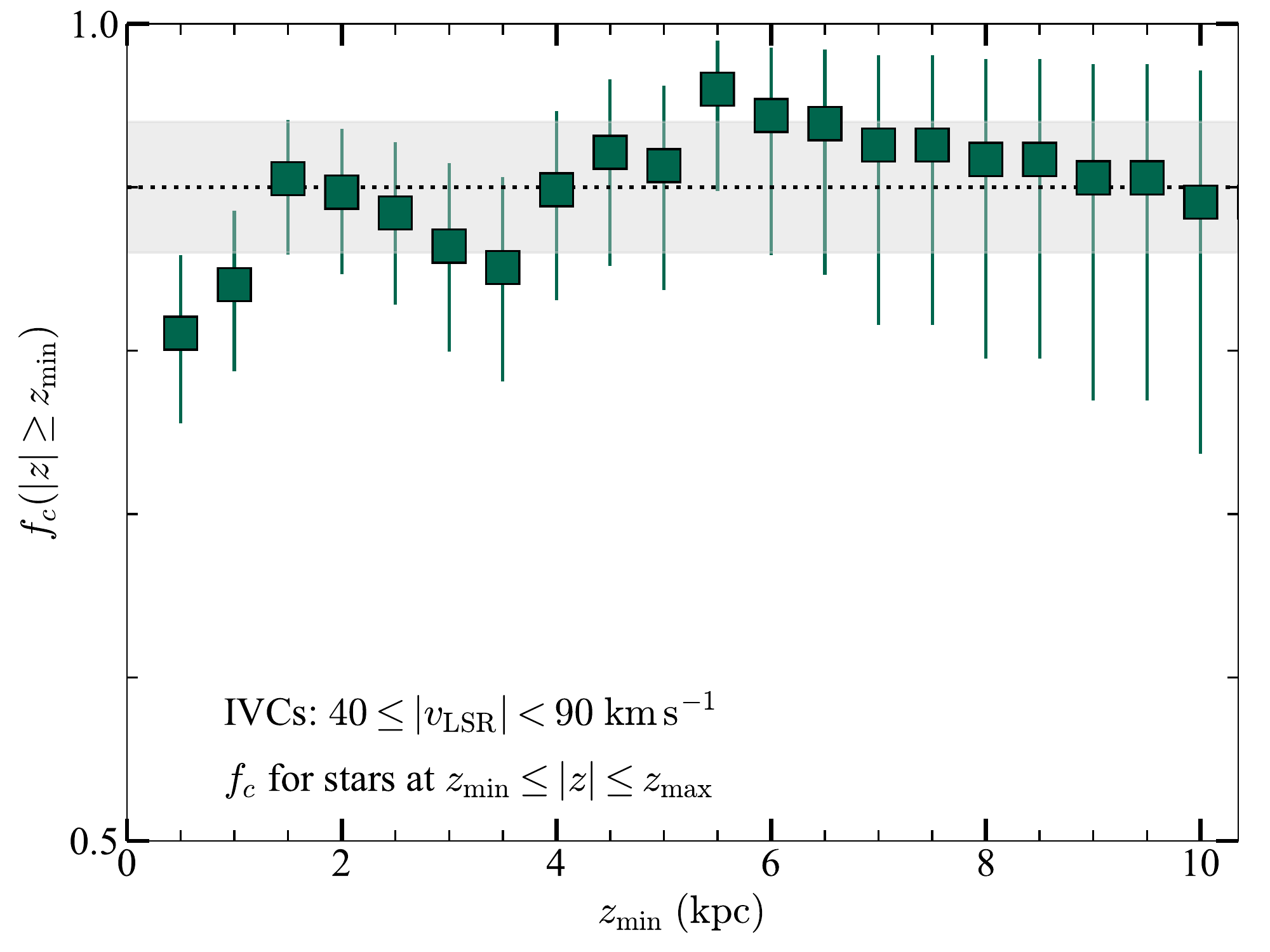}
\caption{Covering factors of the IVCs as a function of the $z$-height determined between $z_{\rm min}  =0.5,...,10$ kpc and $z_{\rm max}= 14.2$ kpc (the maximum $z$-height in our stellar sample). The dotted line (and 1$\sigma$ error bar shown with the gray line) shows the mean covering factor of the IVCs determined from the present stellar sample at $|z|\ge 2$ kpc, which is consistent with the mean estimated at larger $z$-heights. The confidence intervals are at the 68\% level.  }
\label{f-fc-ivc}
\end{figure}

Following the methodology in the previous section, we show in Fig.~\ref{f-fc-ivc} the covering factor of the IVCs as a function of $z_{\rm min}$ where $f_c$ is again estimated in several intervals between a given minimum $|z|$-height value and the  maximum $|z|$-height in our stellar sample. In that figure, the horizontal dashed line and gray area show the mean covering factor for stars at $ |z| \ge 2$ kpc, $\langle f_c \rangle = 0.90 \pm 0.04$, which is consistent with the mean estimated in the various ranges $ |z| \ga 1, 2,3,4$ kpc within the $1\sigma$ dispersion. This is a factor $\sim 2$ larger than found for the \caii\ detection towards QSOs \citep{benbekhti12}. This is not too surprising since \caii\ is a non-dominant ion that probes only neutral and weakly ionised gas (\caiii\ is the dominant ion for calcium, see, e.g., \citealt{sembach00}). 

The behavior of $f_c$ for the IVCs with $z$ is quite different from that of the HVCs with essentially a covering factor that is about constant for all the $z$-values in our sample. Only if $|z|<1$ kpc is there a hint that the covering factor might drop to some degree. The covering factor of the IVCs is also more than twice than that of the \hi\ IVC complexes. Therefore, the IVCs are a population of clouds that is found overwhelmingly in the thick disc of the MW at $|z|\le 1.5$ kpc, which must also be largely ionised. Based on  WHAM observations of the IVC complex K, at least for this IVC, a major source of ionization can be produced by the escape of Lyman continuum radiation from OB stars in the disc and ionizing radiation produced by cooling supernova remnants \citep{haffner01}.

Our IVC analysis has ignored the possible changes of $f_c$ in the southern and northern hemispheres. Contrary to the HVCs, $f_c$ for IVCs is essentially the same in both Galactic hemispheres. There is  no evidence for a dependence of $f_c$ of the IVCs with latitude. Thus our  neglect of hemispherical differences is {\it a posteriori} reasonable. 

\subsection{The vertical distribution of ionised gas in the MW}\label{app-hs}
So far, we have studied how the covering factors of the IVCs and HVCs change with $z$. We can also determine their vertical distributions assuming a model for the vertical gas profile and under the following assumptions. First, we assume that the largely ionised material is associated with an ensemble of similar clouds, and second each absorption feature traces one of these clouds. Finally, the covering fractions of these HVCs towards the most distant stars and QSOs should be similar, which has been proven to be the case in \S\ref{s-fc-hvc}. In this scenario, if a target star at a given $z$-height and Galactic latitude $b$ shows $\mathcal{N}$ features, then $\mathcal{N} \sin b $ is proportional to the cumulative distribution of the ionised material from the midplane up to that $z$. The mean $\avg{ \mathcal{N} \sin b}$ computed over several stars for a given $z$ provides a more precise measurement for this quantity. By using several stars at different $z$, we can re-construct the cumulative cloud number count and fit it with a model for the vertical gas profile. We adopt the same model profile firstly introduced by \citet{oosterloo07} to describe the \hi\ halo of NGC\,891, and later adopted by \citet{marasco19} to model the disk-halo (a.k.a., extra-planar) \hi\ distribution and kinematics in nearby galaxies. This model explicitly consists of a $\sinh(|z|/h)  \times \cosh(|z|/h)^{-2}$ profile where $h$ is a characteristic scale height, featuring a depression for $|z|\rightarrow 0$, a peak at $|z|\simeq 0.88h$, and an exponential decline at larger heights. 

\begin{figure}
\includegraphics[width=\columnwidth]{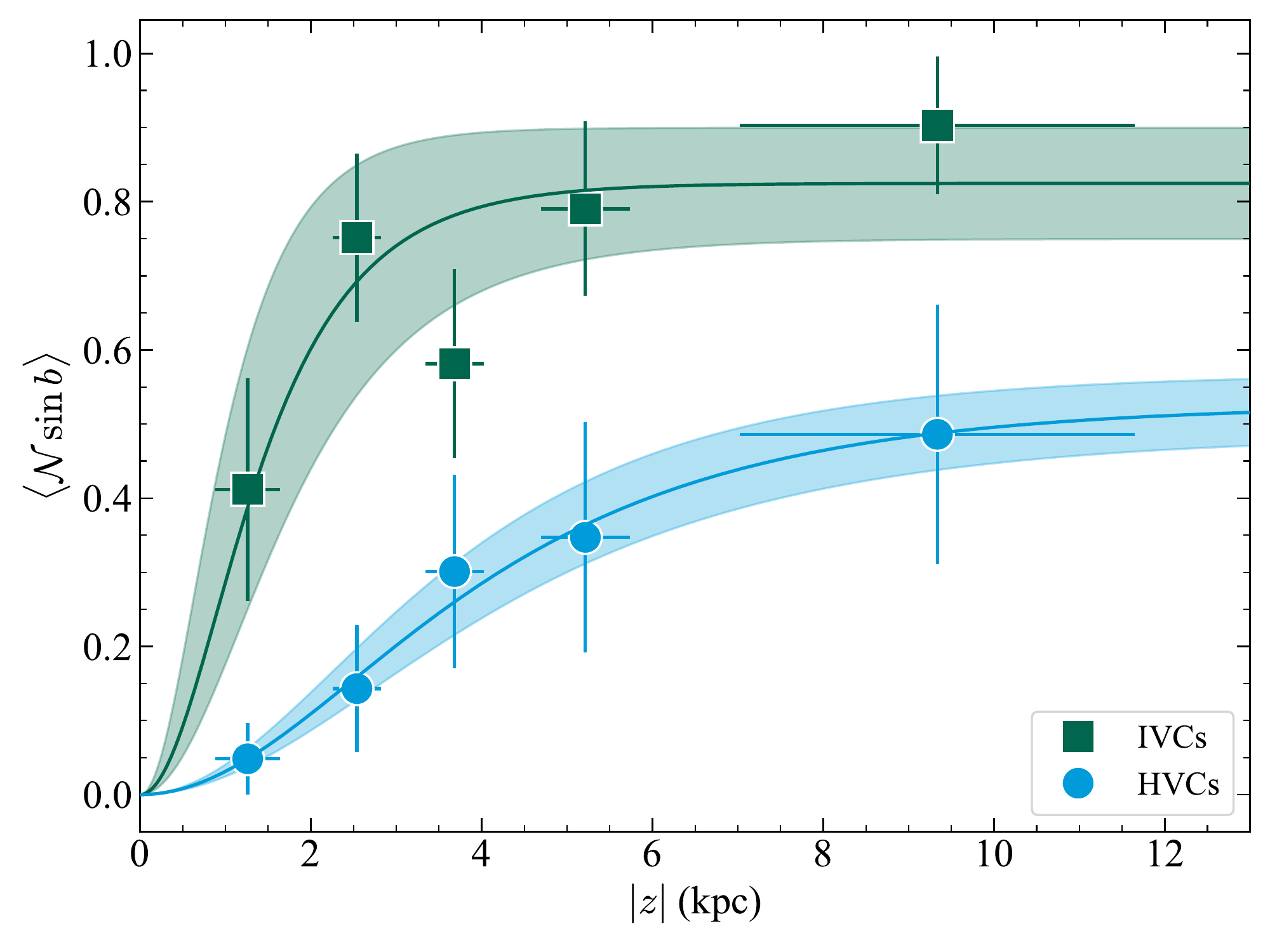}
\caption{Mean number of observed absorbing features corrected for projection effects, $\avg{\mathcal{N} \sin b}$, as a function of $|z|$, for the IVC (squares) and HVCs (circles). The error-bars on the data are determined via bootstrapping. The solid lines shows the prediction for a density profile $\sinh |z|/h \times \cosh (|z|/h)^{-2}$ with a peak at $|z|\simeq 0.88h$  \citep{marasco19} where $h = 1.0 \pm 0.3$ and $2.8 \pm 0.3 $ kpc for the IVCs and HVCs, respectively.}
\label{f-hs}
\end{figure}

In Fig.~\ref{f-hs}, we show these models where the sample is separated between the IVC  and HVC populations, for which we derive scale-heights of $h= 1.0 \pm 0.3$ and $2.8 \pm 0.3$ kpc, respectively. If instead we combine the IVC and HVC populations, then we would find an intermediate scale-height between these two values: $h = 1.4 \pm 0.2 $ kpc.  We note that the scale-heights are of course sensitive to the adopted functional form, but  quite insensitive on our choice for the binning in $z$.


\section{Discussion}\label{s-implications}
In the previous section, we have shown that the covering factor of the HVCs is small (but not zero) at $|z|<2$--3 kpc and the large fraction of the HVCs lies at $3\la |z| \la 7$ kpc above the disc of the MW. For the IVCs, their covering factor does not change beyond $z \simeq 1.5$ kpc, implying that most of the IVCs must be at $z\la 1.5$ kpc.  We also show that the scale-heights are different with $h = 1.0 \pm 0.3$ and $2.8 \pm 0.3$ kpc for the IVCs and HVCs, respectively. HVCs with $90 \la |\vlsr| \la 170$ \km\ therefore predominantly populate the lower Galactic halo (or disc-halo interface), while IVCs largely reside in the thick disc of the MW. Below we discuss how the covering factor of the higher-velocity clouds (VHVCs) than HVCs compares with $f_c$ of the HVCs and IVCs and we briefly discuss some implications for the gas flows in the MW halo. For the latter topic, we refer the reader to \citetalias{marasco22}, which will provide a 3D modeling of the kinematics of these clouds and discuss in detail their possible origins.

\subsection{The VHVCs}\label{s-disc-results}
So far, we have ignored the population of VHVCs, i.e., gas clouds with $|\vlsr| >170 $ \km. VHVCs have been detected towards QSOs in both Galactic hemispheres \citep[e.g.,][]{wakker01,sembach03,lehner12,richter17}. As shown by \citetalias{lehner12}, the covering factor of VHVCs in the northern Galactic sky is $f_c = 0.13 \pm 0.03$, while in the southern sky it is substantially larger with $f_c = 0.46 \pm 0.10$. Over the entire Galactic sky, \citetalias{lehner12} derived $f_c = 0.20 \pm 0.04$ for the VHVCs. Most of the VHVCs are associated with the Magellanic Stream at $b < -20\degr$ (explaining their larger covering factor in the southern hemisphere) and its leading arm at $b > +20\degr$. However, as shown from the \hi\ map in Fig.~\ref{f-map-d} of \citet{westmeier18}, some of the other VHVCs are also associated with the anti-center complex and complexes A, C, and H (see also \citealt{wakker04a,richter17}).

Using the present sample of stars, VHVCs are confirmed to be scarce at $|z|\la 10$ kpc since we did not find a single VHVCs in our sample. Using the methodology described in \S\ref{s-comments} and the entire sample of stars at $1 \la |z| \la 14$, the absence of VHVCs leads to a limit on their covering factor of $f_c \le 0.01\,^{+0.02}_{-0.01}$.  For stars with $3\le |z|\la 14$ kpc, this becomes $f_c \le 0.02\,^{+0.03}_{-0.02}$. For stars with only the largest $z$-height at $6 \le |z|\la 14$ kpc, the upper limit on $f_c$ is still very small with $f_c \le 0.04\,^{+0.06}_{-0.03}$. Even if we compare these limits to only the VHVC covering factor towards QSOs in the northern Galactic sky, these limits towards the stars are substantially smaller, which imply that the bulk of the VHVCs must be beyond $d\ga 10$--15 kpc or $|z| \ga 10$--14 kpc. This is not too surprising for the Magellanic Stream and its leading arm thought to be at $d>50$--100 kpc, which are believed to be the most promising candidates for the origin of the bulk of the VHVC population \citep[][]{besla10,donghia16,barger17}. However, some of the other VHVCs are associated with the complexes A, C, H, and the anti-center, which implies that the very high-velocity parts of these HVC complexes may stretch beyond $z\sim 10$--14 kpc. In fact, only detections at $\vlsr > -160$ \km\ were observed to place the distance bracket estimates on complex C ($7.5 <d <11.7$ kpc, \citealt{wakker07,thom08}), implying part of complex C moving at $-220 \la \vlsr \la -170$ \km\ might be indeed at larger distances. 

\begin{figure}
\includegraphics[width=\columnwidth]{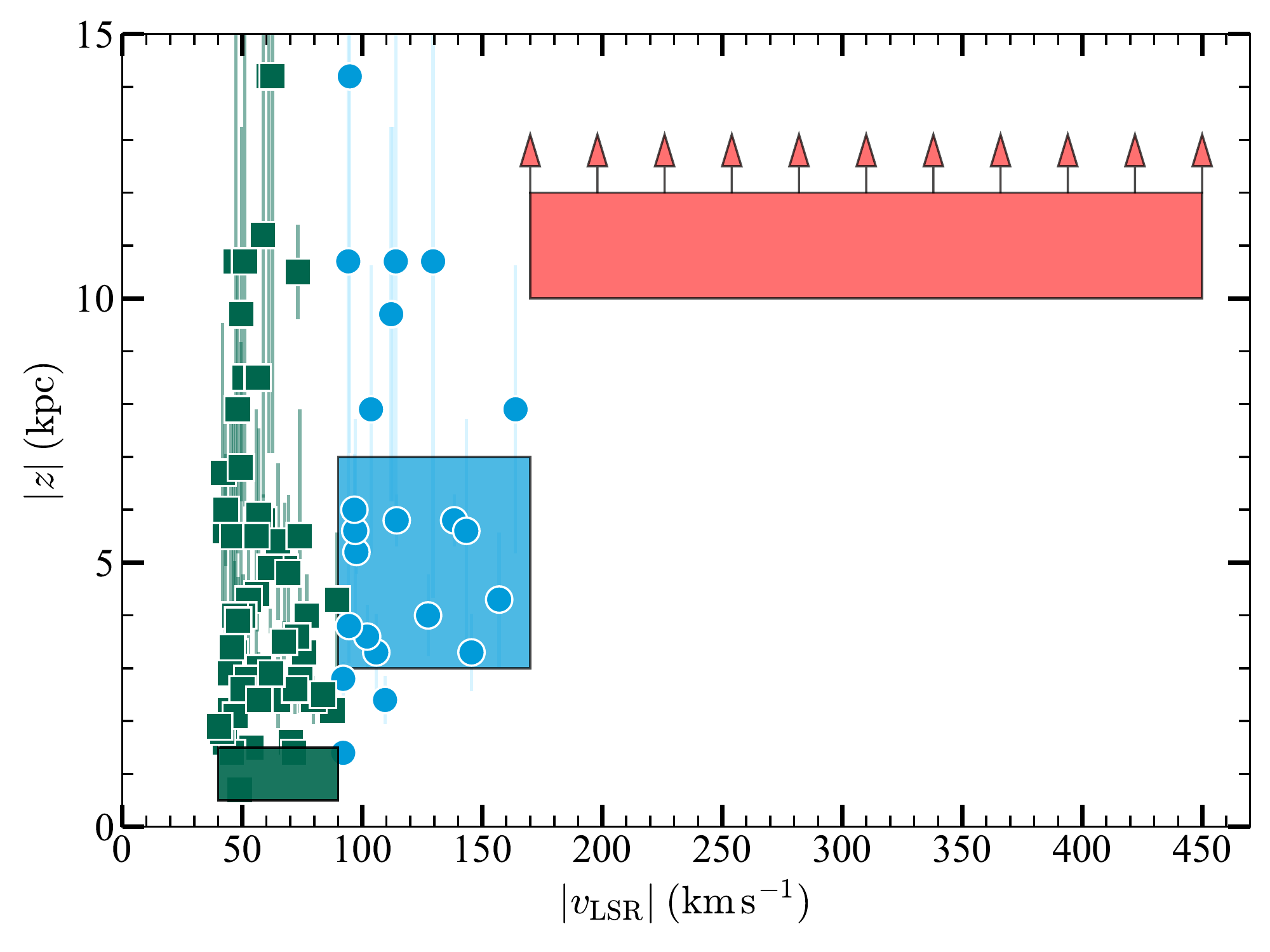}
\caption{The $z$-height of the stars as function of the LSR velocity of the detected IVCs and HVCs (blue circles). For the IVCs and HVCs, $z$ represents an upper limit to the vertical height of the gas cloud. The green and blue rectangles show the most probable locations of the bulk of IVCs and HVCs based on the $f_c$--$z$ analysis. The red rectangle with upward arrows shows the range of velocity of the VHVCs. Since VHVCs are not detected towards stars in our sample, most of the VHVCs are most likely at $|z|\ga 10$--14 kpc.}
\label{f-zh-vc}
\end{figure}

To visualise our findings, we show in Fig.~\ref{f-zh-vc} the vertical height of the stars as a function of the LSR velocity for all the  IVCs and HVCs in our stellar sample. We also show in the green, blue, and red rectangles the most likely $z$ locations of the IVCs, HVCs, and VHVCs, respectively, based on the above conclusions. In summary, VHVCs are found in the halo (circumgalactic medium) of the MW at $|z|\ga 10$ kpc,  HVCs are largely found in the lower halo (or disc-halo interface) at $5 \la |z|\la 10$ kpc (with a smaller fraction at lower $z$), and IVCs are mostly located in the thick disc of the MW at $z\la 1.5$ kpc. 

\subsection{Gas flows in the MW halo}\label{s-disc-interpretation}
Our findings show that clouds with velocities $|\vlsr|>40$ \km\ at $|b|\ga 15\degr$ have a hierarchical distribution of their absolute LSR velocities with $z$: VHVCs are seen only above some $z$-height of 10--15 kpc, IVCs only at small $z$-height of 1--2 kpc, while HVCs are found in the intermediate $z$-height range.  Therefore the absolute LSR velocities of the gas halo clouds decrease with decreasing $z$. We emphasize that this is a property that makes no assumption regarding how the gas clouds is moving in the MW halo (i.e., irrespective if the clouds trace inflows, outflows, or recycled gas). 

We also show that while the covering factor of HVCs decreases with decreasing $z$, 10\% to 40\% of the HVCs are at $z<3$ kpc, implying that HVCs do not only populate the disc-halo interface but can also reach the thick disc of the MW, and eventually be used to fuel new star formation. The high covering factor of IVCs and their location at low $z$ make this population of clouds also prime candidate for the fuel needed to form new stars in the MW disc. On the other hand, VHVCs are not observed in the disc-halo interface or thick disc, and they may end up mostly fueling the MW halo, except if some of the VHVCs are the  higher-velocity precursors of the lower $z$-height HVCs. 

These observational results provide two possible interpretations for the origin of the absorbers studied in this work. The first, a straightforward interpretation  supported by Fig.9 is the ``rain" model by \citet{benjamin97}, according to which drag forces slow down the gas clouds as they fall through the halo. In this scenario, VHVCs in the distant halo of the MW decelerate into HVCs in the disc- halo interface, and then maybe into IVCs in the thick disc of the MW. Thus the IVCs could be remnants of HVCs and VHVCs. 

However, this interpretation runs into several hurdles. First, HVC metallicities range from a few per cent solar to supersolar and most of the HVCs have no evidence of dust depletion \citep[][]{wakker01, richter01,collins03, tripp03,zech08,yao11,tripp12,fox16}. On the other hand, IVCs have a narrower range of metallicities near a solar value and evidence for dust depletion \citep[e.g.,][]{spitzer93,fitzpatrick97,lehner99,richter01,wakker01}. These major differences in metal-enrichment and dust content would require HVCs not only slow down, but strongly mix with the thick disc gas to change their properties into those observed for the IVCs. Second, it is evident from the wide range of metallicities and their association with structures such as the Fermi Bubble \citep{fox15,bordoloi17,ashley20} or the Magellanic Stream \citep[e.g.,][]{putman03,fox14} that HVCs and VHVCs cannot have a single origin.  Although the bulk of the VHVCs probes the Magellanic Stream and its leading arm, they also probe some parts of complex A or C; the latter having properties consistent those expected in a Galactic fountain \citep{fraternali15}. 

In the light of these considerations, the second possible scenario is that HVCs and IVCs originate from a gas cycle triggered by stellar feedback such as the galactic fountain \citep[e.g.,][]{shapiro76,bregman80,fraternali06}. In this framework, clouds are ejected from the disc at different vertical speeds, with the fastest (slowest) clouds reaching higher (lower) heights, producing the HVC and IVC population. Chemical mixing between the metal-rich disc clouds and the metal-poor halo gives rise to a metallicity spread that becomes larger for clouds with longer trajectories, thus IVCs could maintain the metallicity of the disc whereas HVCs would show a higher degree of mixing \citep{marasco12,fraternali15}.

Clearly, the two scenarios described are not mutually exclusive, and it is possible that we are witnessing a combination of both. However, they are characterised by substantially different kinematic signatures: vertical infall should be the dominant kinematic component in the rain model, while rotation should dominate the galactic fountain kinematics. Although the line-of-sight velocity of a single feature is not very informative about the actual 3D kinematics of a single cloud, multiple features sampling different directions in the sky combined with the limits on the distances can be used to assess which model is preferred. In \citetalias{marasco22}, we perform a detailed investigation of the 3D kinematics of the HVCs and IVCs studied in this work, showing that they are best described by a combination of diffuse inflowing and collimated outflowing gas rotating at speed similar to that of the disc, supporting the galactic fountain scenario.

\section{Summary}\label{s-sum}
In this work we have assembled a sample 55 distant halo stars ($d>2$ kpc) that are at vertical heights $|z|\ga 1$ kpc and spread across the Galactic sky to determine how the covering factors of the IVCs and HVCs change with vertical height. Using \hst\ STIS and COS spectra of these 55 stars, we systematically searched for absorption features at high ($|\vlsr| \ge 90$ \km) and intermediate ($40 \le |\vlsr| < 90$ \km) velocities in strong atomic and ionic transitions such as \oi\ $\lambda$1302, \cii\ $\lambda$1334, \civ\ $\lambda\lambda$1548, 1550, \siii\ $\lambda$1190, 1193, 1260, 1304, 1526, \siiii\ $\lambda$1206, \siiv\ $\lambda\lambda$1393, 1402. We summarise the results of our study as follows.

\begin{enumerate}

\item We find 23 HVCs and 58 IVCs in front of our sample of 55 stars at $|z| \ga 1$ kpc. Some of these stars are serendipitously found in the direction of known \hi\ 21-cm emission HVCs and IVCs,  confirming or providing new upper limits on their distances. However, about 52\% of the HVCs and 45\% of the IVCs observed in absorption towards the stars are not associated with known \hi\ complexes, implying these clouds must be largely ionised ($\sim$10\% in each category could be the ionized envelopes of some \hi\ complexes).
\item  We show that the IVCs have a covering factor about constant from $z=1.5$ to 14 kpc with $f_c = 0.90 \pm 0.04$. IVCs are therefore confined to about $|z| \la 1.5$ kpc, i.e., they are in the thick disc of the MW.
\item In contrast, we find that the HVCs have a covering factor that decreases with decreasing $z$-height.  Using stars at $|z|< 2$ kpc, we derive a covering factor $f_c = 0.14\,^{+0.11}_{-0.08}$, while for stars at $|z|\ge 7$ kpc, $f_c = 0.60 \pm 0.15$. The latter value is consistent with that derived towards a sample of QSOs, implying that HVCs with $90 \le |\vlsr| \le 170$ \km\ are largely found in the lower halo (disc-halo interface) of the MW. 
\item We do not detect any VHVCs ($|\vlsr | \ga 170$ \km) in our stellar sample, while their covering factor is $f_c =0.20 \pm 0.04$ towards QSOs. This implies these clouds must be generally at $d>10$--15 kpc ($|z|>10$ kpc).
\item  By assuming that each detection originates from a single cloud of ionised gas, we determine the vertical distribution of the clouds by studying the variation of $f_c$ in different bins of $z$. We derive scale-heights of $h= 1.0 \pm 0.3$ kpc and $2.8\pm 0.3$ kpc for the IVCs and HVCs, respectively.
\item Our findings show that clouds with $|\vlsr|>40$ \km\ at $|b|\ga 15\degr$ have a hierarchical distribution of the absolute LSR velocities with $z$ where  $|\vlsr|$ decreases with decreasing $z$-height. This leads to two main interpretations: a ``rain" model where VHVCs flow from the distant halo down to the disc, producing the population of HVCs and IVCs as they get progressively decelerated by the drag force, and a galactic fountain scenario where HVCs and IVCs are its direct consequences. In the latter scenario, VHVCs may mostly feed the MW halo, although it is possible some of them may also decelerate into HVCs as they fall into the lower halo. In view of their locations and covering factors, these findings finally imply that IVCs and HVCs could serve as star-formation fuel in the MW.
\end{enumerate}

\section*{Data availability}
The data underlying this article are available at the Mikulski Archive for Space Telescopes (MAST, https://archive.stsci.edu). The full-reduced COS spectra are available at the HSLA (https://archive.stsci.edu/missions-and-data/hsla). The various tables from this paper will appear at Centre de Donn\'ees astronomiques de Strasbourg (CDS, http://cdsweb.u-strasbg.fr).

\section*{Acknowledgements}
Support for this research was initially provided by NASA through grant HST-GO-12982 from the Space Telescope Science Institute, which is operated by the Association of Universities for Research in Astronomy, Incorporated, under NASA contract NAS5-26555. Based on observations made with the NASA/ESA Hubble Space Telescope, obtained from the data archive at the Space Telescope Science Institute. STScI is operated by the Association of Universities for Research in Astronomy, Inc. under NASA contract NAS 5-26555. This work has made use of data from the European Space Agency (ESA) mission {\it Gaia} (\url{https://www.cosmos.esa.int/gaia}), processed by the {\it Gaia} Data Processing and Analysis Consortium (DPAC, \url{https://www.cosmos.esa.int/web/gaia/dpac/consortium}). Funding for the DPAC has been provided by national institutions, in particular the institutions participating in the {\it Gaia} Multilateral Agreement. This research has made use of the SIMBAD database, operated at CDS, Strasbourg, France

\bibliographystyle{mnras}

\clearpage

\end{document}